\documentclass[aps,pra,twocolumn,showpacs,superscriptaddress,floatfix]{revtex4-1}
\usepackage{graphicx,amsmath,amssymb}
\usepackage[usenames]{color}
\usepackage[dvipsnames]{xcolor}
\usepackage[colorlinks=true,linkcolor=blue,anchorcolor=blue,citecolor=blue,urlcolor=blue]{hyperref}
\begin{document}

\title{Asymmetric polaron picture for the quantum Rabi model}

\author{Feng Qiao}
\affiliation{School of Physical Science and Technology, Lanzhou University, Lanzhou 730000, China}
\affiliation{Key Laboratory for Quantum Theory and Applications of MoE, Lanzhou Center for Theoretical Physics, Lanzhou University, Lanzhou 730000, China}

\author{Qiu-Yi Chen}
\affiliation{School of Physical Science and Technology, Lanzhou University, Lanzhou 730000, China}
\affiliation{Key Laboratory for Quantum Theory and Applications of MoE, Lanzhou Center for Theoretical Physics, Lanzhou University, Lanzhou 730000, China}

\author{Zu-Jian Ying}
\email{yingzj@lzu.edu.cn}
\affiliation{School of Physical Science and Technology, Lanzhou University, Lanzhou 730000, China}
\affiliation{Key Laboratory for Quantum Theory and Applications of MoE, Lanzhou Center for Theoretical Physics, Lanzhou University, Lanzhou 730000, China}

\begin{abstract}
The experimental access to ultra-strong couplings in light-matter interactions has made the
quantum phase transition (QPT) in the quantum Rabi model practically relevant, while the physics of the QPT has not yet been fully explored.
The polaron picture is a method capable of analyzing in the entire coupling regime and extracting the essential
physics behind the QPT. However, the asymmetric deformation of polarons is missing in the current polaron picture. In the present work we propose an improved variational method
in asymmetric polaron picture (APP). Our APP not only increases the method accuracy but also reveals more underlying physics concerning the QPT. We find that in the ground state both the polarons and antipolarons are asymmetrically deformed to a large extent, which leads to a richer phase diagram. We also analyze the first excited state in
which we unveil an asymmetry direction reversal for the polarons and an attraction/replusion transition differently from the ground state. Finally, we apply the APP in quantum Fisher information analysis and critical coupling extraction, the improvements indicate that the polaron asymmetry makes a considerable contribution to the quantum resource in quantum metrology and plays an unnegligible role in the QPT.  Our
results and mechanism clarifications expose more subtle energy competitions and abundant physics, and the method potentially might have broader applications in light-matter interactions.
\end{abstract}
\pacs{ }
\maketitle


\section{Introduction}\label{Sect-Intro}

The past two decades have seen the fast growing of the field of light-matter
interaction in the frontiers of modern qunatum physics. Both the theoretical
progesses~\cite{Braak2011,Solano2011,Boite2020,Liu2021AQT} and experimental
adavances~\cite{Ciuti2005EarlyUSC,Aji2009EarlyUSC,Diaz2019RevModPhy,Kockum2019NRP,Wallraff2004,Gunter2009, Niemczyk2010,Peropadre2010,FornDiaz2017,Forn-Diaz2010,Scalari2012,Xiang2013,Yoshihara2017NatPhys,Kockum2017,Bayer2017DeepStrong,Ulstrong-JC-3-Adam-2019,Qin2024PhysRep}
have brought much attention to the quantum Rabi model (QRM)~\cite{rabi1936,Rabi-Braak,Eckle-Book-Models} which is a most
fundamental model of light-matter interactions. Indeed, the milestone work~\cite{Braak2011}
revealing the integrability of the QRM has induced a massive dialogue~\cite{Solano2011}
between mathematics and physics~\cite{Braak2011,Solano2011,Boite2020,Liu2021AQT,
Ashhab2013,Ying2015,Liu2021AQT,Hwang2015PRL,Hwang2016PRL,Irish2017,
Ying-g2hz-QFI-2024,*Ying-g2hz-QFI-2024-Cover,Ying-g1g2hz-QFI-2025,Ying2025g2A4,*Ying2025g2A4-Cover,Ying-g2Stark-QFI-2025,
LiuM2017PRL,Ying-2018-arxiv,Ying2020-nonlinear-bias,Ying-2021-AQT,*Ying-2021-AQT-Cover,
Ying-gapped-top,
Ying-Stark-top,*Ying-Stark-top-Cover,
Ying-Spin-Winding,*Ying-Spin-Winding-Cover,
Ying-JCwinding,Ying-Topo-JC-nonHermitian,*Ying-Topo-JC-nonHermitian-Cover,Ying-Topo-JC-nonHermitian-Fisher,*Ying-Topo-JC-nonHermitian-Fisher-Cover,Ying-gC-by-QFI-2024, Grimaudo2022q2QPT,Grimaudo2023-Entropy,Grimaudo2024PRR,Zhu2024PRL,DeepStrong-JC-Huang-2024,PengJie2019,Padilla2022,Gao2022Rabi-dimer,GaoXL2025SPT,QiuYi2025gA2,
Garbe2020,Montenegro2021-Metrology,Chu2021-Metrology,Garbe2021-Metrology,Ilias2022-Metrology, Ying2022-Metrology,YangZheng2023SciChina,Gietka2023PRL-Squeezing,Hotter2024-Metrology,Alushi2024PRL,Mukhopadhyay2024PRL,Mihailescuy2024,
Ying-Topo-JC-nonHermitian-Fisher,*Ying-Topo-JC-nonHermitian-Fisher-Cover,Ying-g2hz-QFI-2024,*Ying-g2hz-QFI-2024-Cover,
Ying-g1g2hz-QFI-2025,Ying2025g2A4,*Ying2025g2A4-Cover,Ying-g2Stark-QFI-2025,Gietka2025PRL100802,Mihailescu2025CQMtutorial,QiuYi2025gA2,
Bera2014Polaron,
CongLei2017,CongLei2019,ChenQH2012,
Braak2019Symmetry,
Wolf2012,FelicettiPRL2020,Felicetti2018-mixed-TPP-SPP,Felicetti2015-TwoPhotonProcess,Simone2018,Alushi2023PRX,
Irish2014,Irish-class-quan-corresp,
PRX-Xie-Anistropy,Batchelor2015,XieQ-2017JPA,
e-collpase-Garbe-2017,e-collpase-Duan-2016,Rico2020,
Boite2016-Photon-Blockade,Ridolfo2012-Photon-Blockade,Li2020conical,Ma2020Nonlinear,
ZhangYY2016,ZhengHang2017,Zheng2017,Yan2023-AQT,Chen-2021-NC,Liu2015, ChenGang2011-GVM,ChenGang2012,FengMang2013,HiddenSymMangazeev2021,HiddenSymLi2021,HiddenSymBustos2021,Casanova2018npj,JC-Larson2021,Ulstrong-JC-2,Eckle-2017JPA,*Eckle-2017JPA-b,
Stark-Grimsmo2013,Stark-Grimsmo2014,Lu-2018-1,Xie2019-Stark,Stark-Cong2020,Cong2022Peter,Qin-ExpLightMatter-2018,LiPengBo-Magnon-PRL-2024,PengJ2021PRL,Gao2022Rabi-aniso}.
On the other hand, the experimental access
to ultra-strong~\cite{Ciuti2005EarlyUSC,Aji2009EarlyUSC,Diaz2019RevModPhy,Kockum2019NRP,Wallraff2004,Gunter2009,Niemczyk2010, Peropadre2010,FornDiaz2017,Forn-Diaz2010,Scalari2012,Xiang2013,Yoshihara2017NatPhys,Kockum2017,Bayer2017DeepStrong,Qin2024PhysRep,Ulstrong-JC-2}
and deep-strong~\cite{WangYouJQ2023DeepStrong,Yoshihara2017NatPhys,Bayer2017DeepStrong,DeepStrong-JC-Huang-2024}
couplings
has found the indispensable role of the counter-rotating terms~\cite{PRX-Xie-Anistropy} which renders
the QRM to be more important. The QRM is particularly interesting also due
to the fact that it possesses a finite-component quantum phase transition (QPT)~\cite{Ashhab2013,Ying2015,Liu2021AQT,Hwang2015PRL,Hwang2016PRL,Irish2017,
Ying-g2hz-QFI-2024,*Ying-g2hz-QFI-2024-Cover,Ying-g1g2hz-QFI-2025,Ying2025g2A4,*Ying2025g2A4-Cover,Ying-g2Stark-QFI-2025,
LiuM2017PRL,Ying-2018-arxiv,Ying2020-nonlinear-bias,Ying-2021-AQT,*Ying-2021-AQT-Cover,
Ying-gapped-top,
Ying-Stark-top,*Ying-Stark-top-Cover,
Ying-Spin-Winding,*Ying-Spin-Winding-Cover,
Ying-JCwinding,Ying-Topo-JC-nonHermitian,*Ying-Topo-JC-nonHermitian-Cover,Ying-Topo-JC-nonHermitian-Fisher,*Ying-Topo-JC-nonHermitian-Fisher-Cover,Ying-gC-by-QFI-2024, Grimaudo2022q2QPT,Grimaudo2023-Entropy,Grimaudo2024PRR,Zhu2024PRL,DeepStrong-JC-Huang-2024,PengJie2019,Padilla2022,Gao2022Rabi-dimer,GaoXL2025SPT}
as in many-body systems~\cite{LiuM2017PRL,Irish2017}. Such a finite-component QPT
can be practically applied in critical quantum metrology~\cite{Garbe2020,Montenegro2021-Metrology,Chu2021-Metrology,Garbe2021-Metrology,Ilias2022-Metrology, Ying2022-Metrology,YangZheng2023SciChina,Gietka2023PRL-Squeezing,Hotter2024-Metrology,Alushi2024PRL,Mukhopadhyay2024PRL,Mihailescuy2024,
Ying-Topo-JC-nonHermitian-Fisher,*Ying-Topo-JC-nonHermitian-Fisher-Cover,Ying-g2hz-QFI-2024,*Ying-g2hz-QFI-2024-Cover,
Ying-g1g2hz-QFI-2025,Ying2025g2A4,*Ying2025g2A4-Cover,Ying-g2Stark-QFI-2025,Gietka2025PRL100802,Mihailescu2025CQMtutorial,QiuYi2025gA2}.

In the afore-mentioned dialogue
between mathematics and physics~\cite{Solano2011},  various approaches and methods~\cite{Boite2020} have been developed for investigations on the QRM and its extensions, such as  exact solution in Bargmann-space representation~\cite{Braak2011}, Bogoliubov transformation~\cite{ChenQH2012}, exact diagonalization~\cite{Ying2020-nonlinear-bias,Ying-Spin-Winding}, the rotating-wave approximation (RWA)~\cite{JC-model}, the generalized RWA~\cite{Irish2007GRWA}, symmetry analysis~\cite{Braak2019Symmetry,HiddenSymMangazeev2021,HiddenSymLi2021,HiddenSymBustos2021,
Ying-2021-AQT,*Ying-2021-AQT-Cover,Ying-JCwinding,Ying-Topo-JC-nonHermitian-Fisher,*Ying-Topo-JC-nonHermitian-Fisher-Cover,Ying-g2hz-QFI-2024},
the adiabatic
approximation~\cite{Irish-2005-AA}, the generalized variational method~\cite{ChenGang2011-GVM,ChenGang2012}, Schrieffer-Wolff transformation~\cite{Hwang2015PRL,LiuM2017PRL}, the mean-photon-number-dependent variational method~\cite{Liu2015}, mean-field method~\cite{Hwang2016PRL,PengJie2019}, variational
displaced coherent state method~\cite{Irish2014,ZhangYY2016,Hwang2010,Bera2014Polaron},
the polaron picture~\cite{Ying2015,CongLei2019,Ying2020-nonlinear-bias,Ying-gapped-top}, and so on.  These methods have made undeniable contributions in the road to
the numerous findings including integrability~\cite{Braak2011}, hidden
symmetry~\cite{Braak2019Symmetry,HiddenSymMangazeev2021,HiddenSymLi2021,HiddenSymBustos2021},
various patterns of symmetry breaking~\cite{Ying2020-nonlinear-bias,Ying-2018-arxiv,Ying-2021-AQT},
finite-component QPTs~\cite{Ashhab2013,Ying2015,Liu2021AQT,Hwang2015PRL,Hwang2016PRL,Irish2017,
Ying-g2hz-QFI-2024,*Ying-g2hz-QFI-2024-Cover,Ying-g1g2hz-QFI-2025,Ying2025g2A4,*Ying2025g2A4-Cover,Ying-g2Stark-QFI-2025,
LiuM2017PRL,Ying-2018-arxiv,Ying2020-nonlinear-bias,Ying-2021-AQT,*Ying-2021-AQT-Cover,
Ying-gapped-top,
Ying-Stark-top,*Ying-Stark-top-Cover,
Ying-Spin-Winding,*Ying-Spin-Winding-Cover,
Ying-JCwinding,Ying-Topo-JC-nonHermitian,*Ying-Topo-JC-nonHermitian-Cover,Ying-Topo-JC-nonHermitian-Fisher,*Ying-Topo-JC-nonHermitian-Fisher-Cover,Ying-gC-by-QFI-2024, Grimaudo2022q2QPT,Grimaudo2023-Entropy,Grimaudo2024PRR,Zhu2024PRL,DeepStrong-JC-Huang-2024,PengJie2019,Padilla2022,Gao2022Rabi-dimer,GaoXL2025SPT},
multicriticalities and multiple points~\cite{Ying2020-nonlinear-bias,Ying-2021-AQT,Ying-gapped-top,Ying-Stark-top},
universality classification~\cite{Hwang2015PRL,LiuM2017PRL,Irish2017,Ying-2021-AQT,Ying-Stark-top,Ying-Topo-JC-nonHermitian-Fisher,*Ying-Topo-JC-nonHermitian-Fisher-Cover},
spectral collapse~\cite{Felicetti2015-TwoPhotonProcess,e-collpase-Garbe-2017,e-collpase-Duan-2016,CongLei2019,Rico2020} and stablization~\cite{Ying2025g2A4,*Ying2025g2A4-Cover},
photon blockade effect~\cite{Boite2016-Photon-Blockade,Ridolfo2012-Photon-Blockade},
spectral conical intersections~\cite{Li2020conical},
classical-quantum correspondence~\cite{Irish-class-quan-corresp},
single-qubit conventional and unconventional topological phase transitions~\cite{Ying-2021-AQT,Ying-gapped-top,Ying-Stark-top,Ying-Spin-Winding,Ying-JCwinding},
coexistence and simultaneous occurrence of Landau-class and topological-class phase transitions~\cite{Ying-2021-AQT,Ying-Stark-top,Ying-JCwinding,Ying-Topo-JC-nonHermitian-Fisher,*Ying-Topo-JC-nonHermitian-Fisher-Cover}, robust topological feature against nonhermiticity~\cite{Ying-Topo-JC-nonHermitian,*Ying-Topo-JC-nonHermitian-Cover,Ying-Topo-JC-nonHermitian-Fisher,*Ying-Topo-JC-nonHermitian-Fisher-Cover},
squeezing and critical resources for quantum metrology~\cite{Garbe2020,Montenegro2021-Metrology,Chu2021-Metrology,Garbe2021-Metrology,Ilias2022-Metrology, Ying2022-Metrology,YangZheng2023SciChina,Gietka2023PRL-Squeezing,Hotter2024-Metrology,Alushi2024PRL,Mukhopadhyay2024PRL,Mihailescuy2024,
Ying-Topo-JC-nonHermitian-Fisher,*Ying-Topo-JC-nonHermitian-Fisher-Cover,Ying-g2hz-QFI-2024,*Ying-g2hz-QFI-2024-Cover,
Ying-g1g2hz-QFI-2025,Ying2025g2A4,*Ying2025g2A4-Cover,Ying-g2Stark-QFI-2025,Gietka2025PRL100802,Mihailescu2025CQMtutorial,QiuYi2025gA2},
and so forth.

Essentially, many of these achievements are in a direct or indirect connection with the QPT originally found in the QRM~\cite{Ashhab2013,Ying2015,Liu2021AQT,Hwang2015PRL,Hwang2016PRL,Irish2017,
Ying-g2hz-QFI-2024,*Ying-g2hz-QFI-2024-Cover,Ying-g1g2hz-QFI-2025,Ying2025g2A4,*Ying2025g2A4-Cover,Ying-g2Stark-QFI-2025,
LiuM2017PRL,Ying-2018-arxiv,Ying2020-nonlinear-bias,Ying-2021-AQT,*Ying-2021-AQT-Cover,
Ying-gapped-top,
Ying-Stark-top,*Ying-Stark-top-Cover,
Ying-Spin-Winding,*Ying-Spin-Winding-Cover,
Ying-JCwinding,Ying-Topo-JC-nonHermitian,*Ying-Topo-JC-nonHermitian-Cover,Ying-Topo-JC-nonHermitian-Fisher,*Ying-Topo-JC-nonHermitian-Fisher-Cover,Ying-gC-by-QFI-2024, Grimaudo2022q2QPT,Grimaudo2023-Entropy,Grimaudo2024PRR,Zhu2024PRL,DeepStrong-JC-Huang-2024,PengJie2019,Padilla2022,Gao2022Rabi-dimer,GaoXL2025SPT}.  An effective method in the investigation of the QPT in the QRM is the variational method of polaron picture, which is valid with both high accuracy and clear physical picture for the entire coupling regime, thus facilitating the analysis on the transition~\cite{Ying2015}. The polaron picture has also been effectively applied in extracting various patterns of symmetry breaking~\cite{Ying2020-nonlinear-bias} in non-linear couplings~\cite{Ying2020-nonlinear-bias,CongLei2019,Ying-2018-arxiv}, understanding the topological transitions~\cite{Ying-2021-AQT,Ying-gapped-top,Ying-Stark-top}, and comprehending the interference fringes in the Wigner function and the hidden squeezing transition~\cite{Ying-gapped-top}. The polaron picture includes polarons with displacement and frequency renormalization which capture the main physics in the QPT~\cite{Liu2021AQT,Ying2015,Ying2020-nonlinear-bias,CongLei2019,Ying-2018-arxiv,Ying-2021-AQT,Ying-gapped-top,
Ying-Stark-top,CongLei2017,CongLei2019,Ying-2018-arxiv,Ying-gC-by-QFI-2024,
Ying-g2hz-QFI-2024,*Ying-g2hz-QFI-2024-Cover,
Ying-g1g2hz-QFI-2025,Ying-g2Stark-QFI-2025}.  However, the considered polarons are symmetric, while the subtle energy competitions may deform the polarons and lead to asymmetry. A detailed analysis on such a missing physics is lacking.

In this work, we propose a variational method of asymmetric polaron picture
which incorporates the missing polaron asymmetry degree of freedom. The method
not only leads to a higher accuracy than the previous symmetric polaron
picture but also reveals more physics that was not captured in the symmetric
polaron picture. Indeed, we find that in the ground state both the polarons
and antipolarons are asymmetrically deformed to a large extent in the entire
coupling regime, which reflects the attracting of the polaron and
antipolaron in the ground state. Such an asymmetry reaches a maximum around
the QPT. We also find an emerging transition of imbalance reversal for the
asymmetry extent of the polarons and the antipolarons. We also analyze the
first excited state in which we find asymmetry direction transition for the
polarons, which reflects a repelling tendency of the polarons and the
antipolarons oppositely to the ground state. Finally, we apply the method for quantum Fisher information
analysis and critical coupling extraction, with the improvements over the symmetric polaron picture.
Our results reveal more underlying physics from the subtle energy competitions in light-matter
interactions, with potential for broader applications.

The paper is organized as follows.
Section~\ref{Sect-Model} introduces the QRM with the transform to the position space.
Section~\ref{Sect-Method} presents our variational method in the asymmetric polaron picture and shows the improved accuracy.
Section~\ref{Sect-GS} shows the features, transitions and the phase diagram of the ground state extracted by the asymmetric polaron picture.
Section~\ref{Sect-Excited-State} analyzes the behavior of the first excited state and the phase diagram with additional transitions of polaron asymmetry.
Section~\ref{Sect-Fq-gc} applies the method and shows the improvements in quantum Fisher information and critical couplings.
Section~\ref{Sect-Conclusions} summarizes the conclusions.

\section{Model and symmetry}

\label{Sect-Model}

The QRM~\cite{rabi1936,Rabi-Braak,Eckle-Book-Models} is
a most fundamental model for light-matter interactions, describing the
coupling of a two-level system (qubit) with a single-mode light field. Under
the rotating-wave approximation it is equivalent to the Jaynes-Cummings
model~\cite{JC-model,JC-Larson2021}. The QRM and its extensions also have a broad relevance,
being a fundamental building block for quantum information and quantum
computation~\cite{Diaz2019RevModPhy,Romero2012,Stassi2020QuComput,Stassi2018,Macri2018},
applied in quantum metrology~\cite{Garbe2020,Montenegro2021-Metrology,Chu2021-Metrology,Garbe2021-Metrology,Ilias2022-Metrology, Ying2022-Metrology,YangZheng2023SciChina,Gietka2023PRL-Squeezing,Hotter2024-Metrology,Alushi2024PRL,Mukhopadhyay2024PRL,Mihailescuy2024,
Ying-Topo-JC-nonHermitian-Fisher,*Ying-Topo-JC-nonHermitian-Fisher-Cover,Ying-g2hz-QFI-2024,*Ying-g2hz-QFI-2024-Cover,
Ying-g1g2hz-QFI-2025,Ying2025g2A4,*Ying2025g2A4-Cover,Ying-g2Stark-QFI-2025,Gietka2025PRL100802,Mihailescu2025CQMtutorial,QiuYi2025gA2},
bridged to many-body systems~\cite{LiuM2017PRL,Irish2017},
connected to models in condense matter~\cite{Kockum2019NRP} and couplings in nanowires~\cite{Nagasawa2013Rings,Ying2016Ellipse,Ying2017curvedSC,Ying2020PRR,Gentile2022NatElec}
and cold atoms~\cite{LinRashbaBECExp2013Review,LinRashbaBECExp2011,LiuYing02025exoticSOC2Ring,*LiuYing02025exoticSOC2Ring-Cover,LiuYing02025KaleidoscopeDDI}. The Hamiltonian
of the standard QRM reads
\begin{equation}
H_{R}=\omega {\hat{a}^{\dagger }\hat{a}}+\frac{\Omega }{2}\sigma
_{x}+g\sigma _{z}(\hat{a}^{\dagger }+\hat{a})  \label{eq:rabi}
\end{equation}%
where $\hat{a}^{\dagger }(\hat{a})$ is the bosonic creation (annihilation)
operator for the light field with frequency $\omega $. The two-level system
is represented by the Pauli matrices $\sigma _{x,y,z}$ with $\Omega $ being
the level splitting. The coupling strength is denoted by $g$. In recent
years, ultra-strong~\cite{Ciuti2005EarlyUSC,Aji2009EarlyUSC,Diaz2019RevModPhy,Kockum2019NRP,Wallraff2004,Gunter2009,Niemczyk2010, Peropadre2010,FornDiaz2017,Forn-Diaz2010,Scalari2012,Xiang2013,Yoshihara2017NatPhys,Kockum2017,Bayer2017DeepStrong,Qin2024PhysRep}
and deep-strong~\cite{WangYouJQ2023DeepStrong,Yoshihara2017NatPhys,Bayer2017DeepStrong} couplings
have been realized in superconducting circuit systems. Here, the adopted
spin notation as in Refs.\cite{Irish2014,Ying2015} which conveniently represents
the two flux states in the flux-qubit circuit systems~\cite{flux-qubit-Mooij-1999} by $\sigma _{z}=\pm $. One can exchange $\sigma _{x}$
and $\sigma _{z}$ to retrieve the conventional form of the QRM by a spin
rotation around the axis $\vec{x}+\vec{z}$.

By the transform $\hat{a}^{\dagger }=(\hat{x}-i\hat{p})/\sqrt{2}$, $\hat{a}=(%
\hat{x}+i\hat{p})/\sqrt{2}$, where $\hat{x}=x$ and $\hat{p}=-i\frac{\partial
}{\partial x},$ the Hamiltonian can be mapped onto the position space and
rewritten as
\begin{equation}
H=\sum_{\sigma _{z}=\pm }\left( h^{\sigma _{z}}\left\vert \sigma
_{z}\right\rangle \left\langle \sigma _{z}\right\vert +\frac{\Omega }{2}%
\left\vert \sigma _{z}\right\rangle \left\langle \overline{\sigma }%
_{z}\right\vert \right) +\varepsilon _{0}.  \label{eq:rabi rewrite}
\end{equation}%
Here $\overline{\sigma }_{z}=-\sigma _{z}$ and $h^{\pm }=\frac{1}{2}\omega (%
\hat{p}^{2}+v_{\pm })$ is an effective single-particle energy in a
spin-dependent harmonic-oscillator potential $v_{\pm }=(\hat{x}\pm g^{\prime
})^{2}$ with a potential displacement $g^{\prime }=\sqrt{2}g/\omega $, while
$\varepsilon _{0}=-\frac{1}{2}\omega (g^{\prime 2}+1)$ is a constant energy.
Now on the $\sigma _{z}$ basis $\Omega $ becomes the strength of tunneling
in the position space or spin flipping in the spin space~\cite%
{Irish2014,Ying2015}.

The model has parity symmetry, $[{\cal P},H]=0$, where the parity operator $%
{\cal P}=\sigma _{x}(-1)^{\hat{a}^{\dagger }\hat{a}}$ contains simultaneous
spin reversal $\pm \rightarrow \mp $ and space inversion $x\rightarrow -x$~%
\cite{Ying2020-nonlinear-bias}. Therefore, the wave function takes the form $%
\left\vert \Psi \right\rangle =\frac{1}{\sqrt{2}}(\left\vert \psi
_{+}\right\rangle \otimes \left\vert \uparrow \right\rangle +\left\vert \psi
_{-}\right\rangle \otimes \left\vert \downarrow \right\rangle )$ where%
\begin{equation}
\psi _{+}\left( x\right) =\psi (x),\quad \psi _{-}\left( x\right) =P\psi
(-x).   \label{psi-up-down}
\end{equation}%
Here the spin-up $\uparrow $ (spin-down $\downarrow $) label is an
alternative representation for state $\sigma _{z}=+$ ($-$) and $P=1$ $(-1)$
corresponds to positive (negative) parity. The ground state has a negative
parity which gains a negative tunneling energy, being more favorable to
lower the energy.

\begin{figure}[t]
	\centering
	\includegraphics[width=1.0\columnwidth]{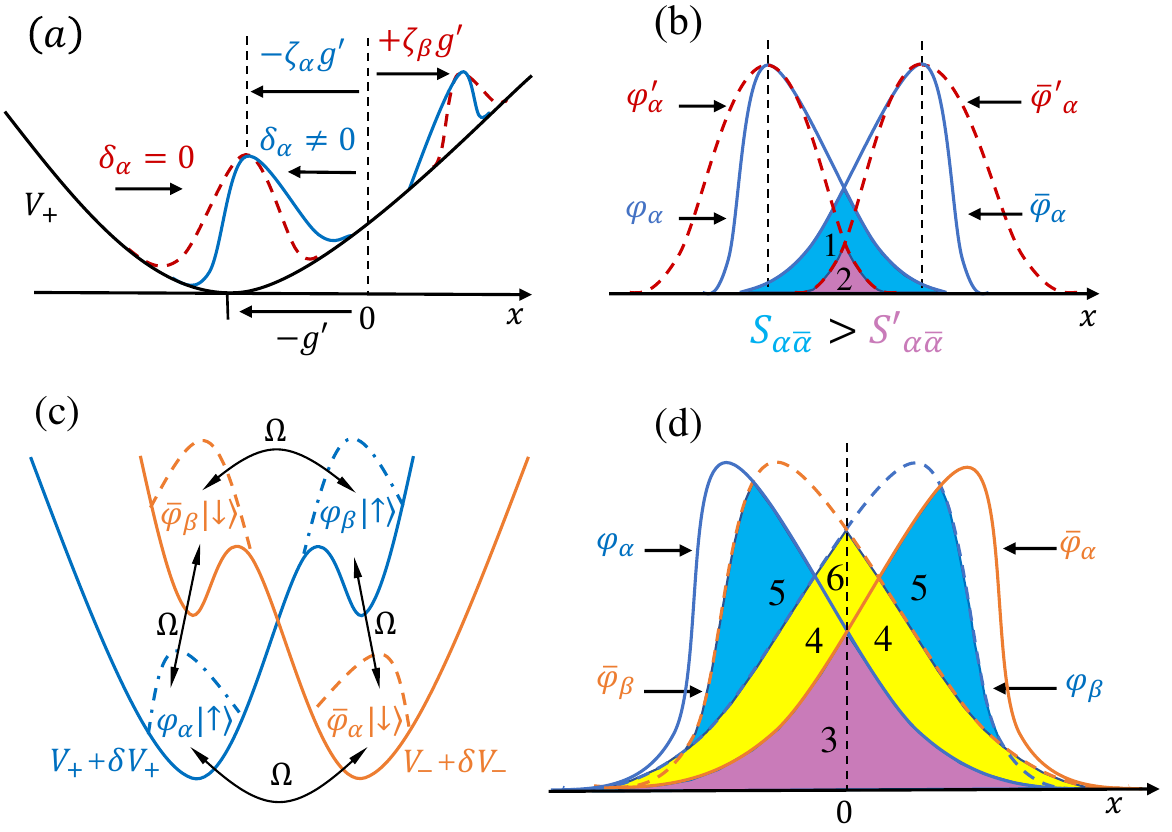}
	\caption{
{\it Schematic diagram for asymmetric polaron picture.} (a) Polarons labeled by $\alpha$  (antipolaron labeled by $\beta$) with displacement $-\zeta_\alpha g ^\prime$ ($+\zeta_\beta g ^\prime$) renormalized from the potential displacement $-g ^\prime$ of $v_{+}$ (parabolic line). Symmetric (Asymmetric) polaron and antipolaron in dashed (solid) lines have an asymmetric factor $\delta_i=0$ ($\delta_i\neq 0$) where $i=\alpha,\beta$.  (b) Enlarged overlap for asymmetric wave packets (solid) relative to symmetric ones (dashed). The overline over $\varphi$ denotes spin-down (- or $\downarrow$) component. (c) Four channels of tunneling between spin-up (+ or $\uparrow$) and spin-down components of polaron and antipolarons accommodated in effective potentials $v_{\pm}+\delta v_{\pm}$. (d) Different overlaps among farther asymmetric polarons and closer asymmetric antipolarons due to potential difference in (a).}
	\label{fig1}
\end{figure}

\section{Asymmetric polaron picture\label{Sect-Method}}

We shall first analyze the physical picture in the QPT of the QRM. With the extracted physical ingredients that play roles in the QPT, we propose the variational method in
asymmetric polaron picture.

\subsection{Physics in asymmetric polaron picture}\label{Sect-polaron-picture}

{\it Potential displacements}: In the absence of coupling the potential $%
v_{\pm }\rightarrow x^{2}$ is simply a harmonic-oscillator potential and
each spin components of the ground state is a Gaussian wave packet in the
lowest state $\phi _{0}$ of quantum harmonic oscillator. When the coupling
is turned on, the potentials $v_{\pm }$ of the two spin components are
separated and shift in opposite directions with displacements $x_{0}^{\pm
}=\mp g^{\prime }$. In Fig. \ref{fig1}(a) the potential $v_{+}$ is plotted
as the parabolic curve with the displacement represented by the bottom
position at $x_{0}^{+}=-g^{\prime }$ which is shifting to the left side of
the origin (marked by the vertical dashed line at $x=0$). A reflection with
respect to the origin will give the potential $v_{-}.$

{\it Wave-packet displacements and polarons}: Accordingly the wave packet in
each spin component has a tendency to follow the potential in the
displacement. However, the tunneling tends to prevent the displacement of
the wave packets as remaining unmoved around the origin $x=0$ would have a
maximized wave-packet overlap between the two spin components to gain
more negative tunneling energy for the ground state. A larger coupling will lead to larger
potential displacement thus a higher potential cost for the wave packets to
stay around the origin, and finally the wave packets will also move away
from the origin but with a discount of the potential displacement,
$\zeta _{\alpha }g^{\prime }$, to partially keep the tunneling energy. The
discounted wave-packet displacement is shown in Fig.~\ref{fig1}(a) by the
other vertical line at $x_{\alpha }^{+}=-\zeta _{\alpha }g^{\prime }$ which
marks the peak position of the moving wave packet in the dashed curve.\
Correspondingly the wave packet in the spin-down component is located around
$x_{\alpha }^{-}=-\zeta _{\alpha }g^{\prime }$ due to the parity symmetry.
These wave packets with displacements in the same direction of the
potentials represent the so-called polarons~\cite{Ying2015,Bera2014Polaron}. The term
of ``polaron" is borrowed from condensed matter in the sense
that a displaced spin state is accompanied with a finite number of photons
as in a coherent state, which is analogous to the electron surrounded by a
phonon cloud due to the atom deviation from the lattice site. Here we denote
the polarons by $\varphi _{\alpha }$ for spin-up component and $\overline{%
\varphi }_{\alpha }(x)$ for spin-down component.

{\it Induced wave packets and antipolarons}: On the other hand, the
tunneling will induce another kind of wave packets with the displacements $x_{\beta
}^{\pm }=\pm \zeta _{\beta }g^{\prime }$ in the two spin components but on the other sides of the origin
that are actually overlapping with the opposite-spin wave packets at
$x_{\alpha }^{\pm }=\mp \zeta _{\alpha }g^{\prime }$. Indeed, the eigen
equation can be rewritten as $\frac{1}{2}\omega (\hat{p}^{2}+v_{\pm }+\delta
v_{\pm })\psi _{\pm }\left( x\right) =E\psi _{\pm }\left( x\right) $ where
\begin{equation}
\delta v_{\pm }=\frac{\Omega }{\omega }\psi _{\mp }(x)/\psi _{\pm }(x)
\end{equation}%
is an induced effective potential in addition to the bare potential $v_{\pm
} $. This induced potential will create a potential well which accommodates a
secondary wave packet, as sketched in Fig. \ref{fig1}(c). We also plot this
secondary wave packet over the bar potential for spin-up component in Fig. %
\ref{fig1}(a) in dashed curve. These secondary wave packets represent
antipolarons in the sense that their displacements are in the opposite
directions of the potentials. Here we denote the antipolarons by $\varphi
_{\beta }(x)$ for the spin-up component and $\overline{\varphi }_{\beta }(x)$
for the spin-down component.

{\it Wave-packet overlap and frequency renormalization}: The polarons and
antipolarons form four channels of tunneling, as indicated by the
double-headed arrows in  Fig.~\ref{fig1}(c), between polarons on the same sides
and between polarons and antipolarons on the opposite sides. The tunneling
strength is decided by the overlaps of the polarons and antipolarons, as
illustrated by the region 2 in Fig. \ref{fig1}(b). To gain more tunneling
energy the wave packet will be extended to expand the overlap region.
Effectively this wave-packet extension can be described by a frequency
renormalization $\omega \rightarrow \xi _{i}\omega $ with $i=\alpha ,\beta $ depending on the
polarons and antipolarons.

{\it Potential height and polaron weight}: As one sees in Fig. \ref{fig1}(a)
the antipolarons have a higher potential than the polarons, which leads to a
weight difference. We denote the weight of the polarons by $\alpha $ and
that of the antipolarons by $\beta $ in the wave-packet superposition. Generally the antipolarons have a
smaller weight than the antipolarons, $\beta <$ $\alpha $.
Counter-intuitively a larger antipolaron weight can still occur as
will be addressed in the discussion later on.

{\it Asymmetric deformation and its origins}: Note both the polarons and the
antipolarons so far are symmetric with respect to their peak position as in
the Gaussian wave packet in $\phi _{0}$. However, the right and left sides
of either a polaron or an antipolaron actually have different potential
values as one may notice in Fig.~\ref{fig1}(a), which would lead to an
imbalance of density distribution on the two sides as sketched by the
solid-line wave-packets in contrast to the dashed-line symmetric ones in Fig.~\ref{fig1}(a).
Moreover, the left-right-side tunneling are dominantly
determined by the overlapping sides of a polaron and an antipolaron, the
wave-packet density on the overlapping sides around the origin has the
leading impact on the tunneling energy relatively to the other sides farther away from
the origin. This side difference in tunneling contribution also results in asymmetric deformation. In Fig.~\ref{fig1}(b)
we compare the asymmetric polarons and antipolarons (solid curves) and
the symmetric ones (dashed curves), which shows the asymmetric ones have a
wider overlapping region (marked by 1) than that of the symmetric ones
(marked by 2).

\subsection{Variational method in asymmetric polaron picture}\label{Sect-Vari-APP}

Based on the above analysis, we propose a trial variational wave function
which takes into account all the key physical ingredients in the asymmetric
polaron picture, via a linear combination of the polaron and the antipolaron
\begin{equation}
\psi (x)=\alpha \varphi _{\alpha }(x)+\beta \varphi _{\beta }(x).
\label{eq:ware function}
\end{equation}%
Here, besides the polaron weights $\alpha $ and $\beta $, the wave packets
$\varphi _{\alpha }(x)$ and $\varphi _{\beta }(x)$ contain the displacement
discount factor $\zeta _{\alpha ,\beta }$, the frequency renormalization
$\xi _{\alpha ,\beta }$, and the asymmetric factor $A_{\alpha ,\beta }$
\begin{eqnarray}
\varphi _{\alpha }(x) &=&(1+A_{\alpha })\phi _{0}(\xi _{\alpha }
,x+\zeta _{\alpha }g^{\prime }), \\
\varphi _{\beta }(x) &=&(1+A_{\beta })\phi _{0}(\xi _{\beta } ,x-\zeta
_{\beta }g^{\prime }),
\end{eqnarray}
where
\begin{equation}
\phi _{0}(x)=\left(
\frac{\xi}{\pi }\right) ^{1/4}\exp [-\frac{1}{2}\xi x^{2}]
\end{equation}
is the ground state of quantum harmonic oscillator with frequency $\xi$.
Although the asymmetric factor may have a more general form, in the
principle of choosing the simplest form to capture the dominant physics, we
take a linear asymmetric order of asymmetric
\begin{equation}
A_{\alpha }=\delta _{\alpha }x_{\alpha }+{\cal O}(x_{\alpha }^{3}),\quad
A_{\beta }=\delta _{\beta }x_{\beta }+{\cal O}(x_{\beta }^{3}),
\label{A-factor}
\end{equation}%
where $x_{\alpha }=x+\zeta _{\alpha }g^{\prime }$ and $x_{\beta }=x-\zeta
_{\beta }g^{\prime }$ denote the distances from the symmetric
polaron and antipolaron positions. The adoption of (\ref{A-factor}) is also based
on the fact that the leading contribution is coming from the region close to
the wave-packet peak, where the linear order plays the key role, while it is
exponentially decaying and vanishing away from the peak. The even-order
terms are symmetric, the effect of which has been encoded in the frequency
renormalization, thus unnecessary to re-include here.

The variational energy can be obtained as
\begin{equation}
E=h_{++}^{+}+\frac{1}{2}P\Omega n_{+-}+\varepsilon _{0}
\end{equation}%
where $h_{++}^{+}=\left\langle \psi \left\vert h^{+}\right\vert \psi
\right\rangle $ and $n_{+-}=\langle \psi |\bar{\psi}\rangle $ with $\bar{\psi}(x)=\psi(-x)$. The
variational parameters $\{\alpha $, $\beta $, $\zeta _{\alpha }$, $\zeta
_{\beta }$, $\xi _{\alpha }$, $\xi _{\beta }$, $\delta _{\alpha }$, $\delta
_{\beta }\}$ are optimized by energy minimization subject to the
normalization condition $\left\langle \psi |\psi \right\rangle =1$. We leave
the detailed explicit expressions in Appendix \ref{Appendix-Vari-E}.

\begin{figure}[t]
	\centering
	\includegraphics[width=1.0\columnwidth]{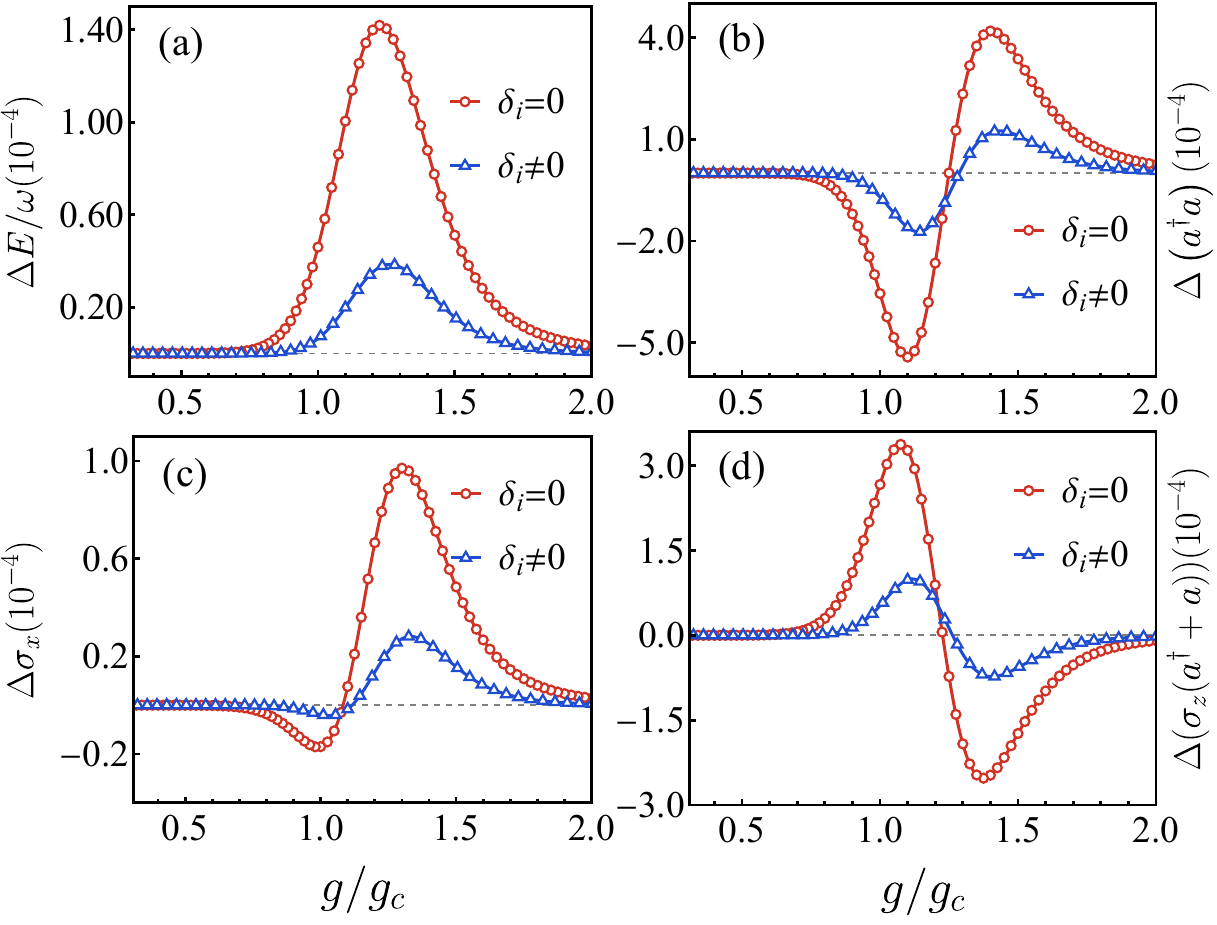}
	\caption{{\it Accuracy improvements in the asymmetric polaron picture.} The discrepancy $\Delta q= q_{var}-q_{\rm ED}$ of variational physical quantities $q_{var}$ from the results of exact diagonalization (ED) for $q$ being (a) the total energy $E$, (b) the photon number $\langle a^\dag a\rangle$, (c) the spin expectation $\langle \sigma_x \rangle$, (d) the coupling correlation $\langle \sigma_z (a^\dag +a)\rangle$.  The circles (triangles) represent the symmetric (asymmetric) case with $\delta _i =0$ ($\delta _i \neq 0$), where $i=\alpha,\beta$,  illustrated for the first excited state at $\omega=0.15\Omega$. }
	\label{Fig-accuracy}
\end{figure}

\subsection{High accuracy of variational method in asymmetric polaron
picture}\label{Sect-Accuracy}

For the QRM the ground state is the lowest state of all eigenstates
including both negative and positive parities, while the first excited state
is the lowest state of all eigenstates with positive parity. The accuracy
for the ground state in the symmetric polaron picture has been addressed for both numerical calculations~\cite{Ying2015,CongLei2017,Ying-gC-by-QFI-2024} and analytic
analysis~\cite{Ying2015,Ying2020-nonlinear-bias,Ying-2018-arxiv,Ying-2021-AQT,Ying-gapped-top,Ying-Stark-top,Ying-gC-by-QFI-2024}.
Here our improved variational method in asymmetric polaron picture is valid for
both the ground state and the first excited state. Fig.~\ref{Fig-accuracy} shows the
discrepancy from the result of exact diagonalization (ED)~\cite{Ying2020-nonlinear-bias,Ying-Spin-Winding} of several physical quantities
of the first excited state, including the energy $E$, the mean photon number
$\langle {\hat{a}^{\dagger }\hat{a}\rangle }$, the tunneling or
spin-flipping strength $\left\langle {\sigma }_{x}\right\rangle $, and
coupling correlation $\langle {\sigma }_{x}({\hat{a}^{\dagger }+\hat{a})\rangle }$.
Here the discrepancy of a quantity $q$ is defined by $\Delta
q=q_{var}-q_{{\rm ED}}$, where $q_{var}$ is variational result and $q_{{\rm ED}}$
is the ED result. In all panels of Fig.~\ref{Fig-accuracy} the
circles are the outcome of symmetric polaron picture ($\delta _{i}=0$) and
the triangles are that of asymmetric polaron picture ($\delta _{i}\neq 0$).
One sees that the discrepancy is substantially reduced in the asymmetric polaron
picture relatively to the results of the symmetric polaron picture. The
accuracy improvement is similar for the ground state. The accuracy
guarantees that an analysis in the asymmetric polaron picture is reliable.
In the next sections we will apply the asymmetric polaron picture to extract
the underlying physics in connection with the PQTs missed by the symmetric polaron
picture.

\section{Polaron asymmetry and transitions in the ground state}
\label{Sect-GS}

Tracking the evolution of the variational parameters in the variation of the
coupling strength will reveal some underlying physics in the ground state. Indeed, we observe
several transition-like features and try to clarify the mechanisms, as
addressed below.

\begin{figure}[t]
	\centering
	\includegraphics[width=1.0\columnwidth]{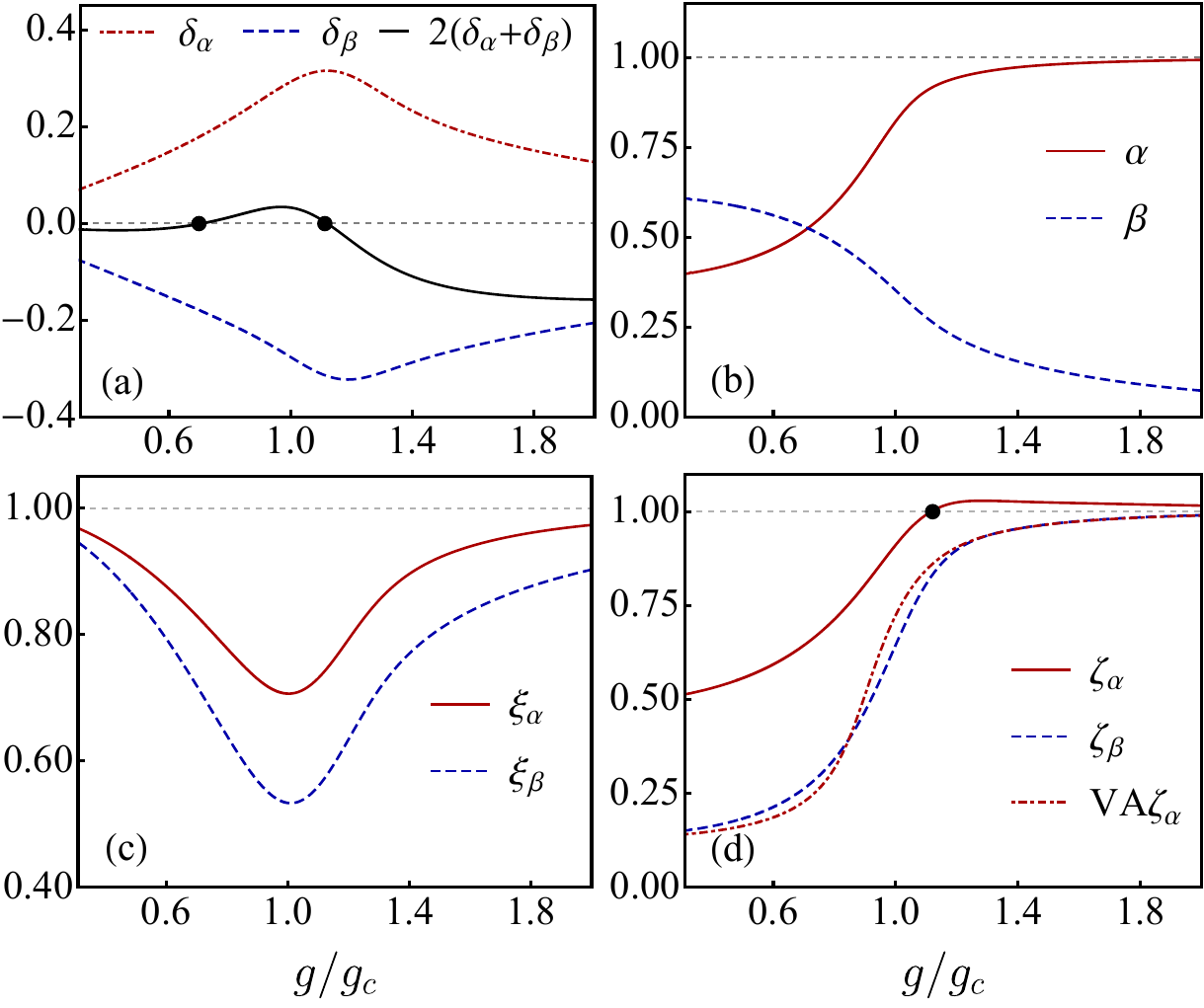}
	\caption{{\it Variational parameters and transitions in the ground state.} Evolution of the variational parameters for (a) asymmetry factors $\delta_\alpha$ (dot-dashed), $\delta_\beta$ (dashed), $\delta_\alpha+\delta_\beta$ (solid), (b) weights of polaron ($\alpha$, solid) and antipolaron ($\beta$, dashed), (c) renormalization factors of polaron ($\xi_\alpha$, solid) and antipolaron ($\xi_\beta$, dashed), (d) displacement renormalization factors of polaron ($\zeta_\alpha$, solid) and antipolaron ($\zeta_\beta$, dashed). The dot-dashed line is the asymmetric polaron displacement $\zeta _{\alpha}+\zeta _{\alpha}^{\delta }$ from Eq.~\eqref{Zata-Asymm}. Here $\omega=0.15\Omega$.}
	\label{Fig-GS-Vars} 	
\end{figure}

\subsection{Transition at maximum frequency renormalization}

The frequency renormalization factors $\xi _{\alpha }$ and $\xi _{\beta }$
are shown in Fig.~\ref{Fig-GS-Vars}(c). One sees that both $\xi _{\alpha }$ and
$\xi _{\beta }$ are smaller than $1$, which means that the distributions of
the wave packets are more extended than the initial wave function $\phi _{0}$
in the absence of coupling. As addressed in Sec.~\ref{Sect-polaron-picture}
such an wave-packet extension enlarges the wave-packet overlap, thus gaining
more tunneling energy which is negative in the ground state. The appearance
of the antipolarons effectively is a precess of wave-packet splitting~\cite{Irish2014,Ying2015}
in which the energy competition between the tunneling
and the potential is going beyond a limit of balance. Around the point of
breaking the balance the wave-packet extension reaches a maximum extent. A
transition-like variation occurs once the wave packets finally split.

The wave-packet splitting occurs around the critical coupling scale~\cite{Ying2015}
\begin{equation}
g_{c}=\sqrt{\omega ^{2}+\sqrt{\omega ^{4}+g_{c0}^{4}}}
\label{Eq-gC}
\end{equation}
which is shifting with finite frequency away from the critical point~\cite{Ashhab2013}
\begin{equation}
g_{c0}= \frac{\sqrt{\omega \Omega }}{2}.
\label{Eq-gC0}
\end{equation}
Eq.~(\ref{Eq-gC0}) can be obtained in a semiclassical consideration in which the spin is kept as the quantum part while the space part takes the classical approximation
by a mass point moving in the classical potential~\cite{Ying2020-nonlinear-bias,Ying-2018-arxiv}. Such a semiclassical approximation is valid in the low-frequency limit, $\omega/\Omega\rightarrow 0$, while Eq.~(\ref{Eq-gC}) works for finite frequencies as the space part is also based on a quantum-mechanics consideration with wave function in wave packets rather than a classical mass point.

Still, Eq.~(\ref{Eq-gC}) only includes the displacement part in the wave-packet splitting,  while further taking into the effect of frequency renormalization leads to a more accurate expression in a fractional-power form~\cite{Ying-gC-by-QFI-2024}
\begin{equation}
g_{c}^{{\rm F}}=g_{c0}\left[ 1+\frac{1}{100\alpha _{{\rm FS}}}(\frac{\omega }{\Omega }%
)^{2/3}-\frac{1}{8}(\frac{\omega }{\Omega })^{4/3}\right] ,  \label{gc2}
\end{equation}%
where $\alpha _{{\rm FS}}=1/137$, 
as extracted from the peak position of the QFI. Expression (\ref{gc2}) is accurate in $\omega /\Omega \in [0,0.5]$ regime, while a higher-order fitting
\begin{equation}
g_{c}^{{\rm F}}=g_{c0}\left[ 1+c_{1}(\frac{\omega }{\Omega }
)^{2/3}+c_{2}(\frac{\omega }{\Omega })^{4/3}+c_{3}(\frac{\omega }{\Omega }
)^{6/3}\right] ,
\label{Eq-gC-by-Fq}
\end{equation}
where $c_{1}=1.3715$, $c_{2}=-0.1311$, $c_{3}=0.0184$ can extend the validity to the entire
frequency regime $\omega /\Omega  \in [0,3]$. Such a fractional-power form comes from the effect of frequency renormalization, in contrast to the less-accurate integer-power form without frequency renormalization~\cite{Ying-gC-by-QFI-2024}.

\subsection{Transition of polaron weight reversal}

Fig.~\ref{Fig-GS-Vars}(b) shows a transition of polaron weight reversal. In a large
coupling $g$ the antipolarons have a smaller weight ($\beta $) than that of
the polarons ($\alpha $) due to the higher antipolaron potential. However, this weight
imbalance can be reversed at a small coupling, especially at low
frequencies, yielding overweighted antipolarons in the sense that they counter-intuitively have a larger weight than the polarons.
In fact, as indicated by the smaller values of $\zeta _{\beta }$
in Fig.~\ref{Fig-GS-Vars}(d), the higher antipolaron potential also leads to a
closer distance between the antipolarons in the two spin components than the
polarons. This yields a larger antipolaron wave-packet overlap than the
polarons, as indicated by Fig.~\ref{fig1}(d) where the antipolaron ovelap
(regions 3,4,6) has an extra part than the that of the poalron (region 4),
which makes a larger weight of antipolarons more favorable to gain more
tunneling energy. At low frequencies, the potential is more flat so that the
overweighted polarons have less potential cost. As a consequence, the
transition of the polaron weight reversal occurs.

\subsection{Considerable asymmetry strength and asymmetry sign\label%
{Sect-large-asym}}

Fig.~\ref{Fig-GS-Vars}(a) illustrates the behavior of the asymmetry parameters
$\delta _{\alpha }$ for the polarons (long-dashed) and $\delta _{\beta }$ for
the antipolarons (short-dashed). The first features we observe are the
asymmetry strength and asymmetry sign. Indeed, the asymmtery is quite
considerable for both the polarons and the antipolarons in the entire
coupling regime, as demonstrated by the finie values of $\delta _{\alpha }$
and $\delta _{\beta }$. This means that the asymmtery effect is fairly
strong and our considerartion in asymmetric polaron picture is necessary.\
Furthermore, $\delta _{\alpha }$ and $\delta _{\beta }$ have oppostite
signs, with the former being positive while the latter being negative,
indicating that $\varphi _{\alpha }(x)$ has more probability on its right
side than its left side while it is reverse for $\varphi _{\beta }(x)$. This
means that it is really the wave-packet sides close to the origin that have
a larger density, which increases the overlaps between left and right wave
packets, as afore-mentioned for Fig.~\ref{fig1}(b). This enlarged overlap
not only gains more negative tunneling energy, but also reduces the
potential energy.

\subsection{Reversals of asymmetry imbalance}\label{Sect-asym-reversal}

A closer look at the competition of $\delta _{\alpha }$ and $\delta _{\beta
} $ brings our attention to the asymmetry imbalance and an underlying
transition of asymmetry imbalance reversal. Indeed, as shown in Fig.~\ref{Fig-GS-Vars}(a) at a large coupling the magnitude of $\delta _{\beta }$ is larger
than that of $\delta _{\alpha }$, which comes from the steeper potential of
the antipolarons than that of the polarons as in Fig.~\ref{fig1}(a).
However, when the coupling strength is reduced, this asymmetry imbalance is
reversed around $g=1.12g_{c}$ (marked by the dot). This reversal arises from
the subtle competitions between the potential difference and the tunneling
energy: (i) First, at a reduced coupling strength the potential of the
antipolarons is less steep, which decreases the potential difference from
the polarons. (ii) Second, the antipolarons and the polarons are getting
closer so that left-right-side tunneling energy becomes more dominant over
the potential gradient difference, due to the enlarged antipolaron-polaron
overlap as sketched by the regions 3,4,6 in Fig.~\ref{fig1}(d). Note here
the left-right polarons have larger distance than the left-right
antipolarons so that the polarons need a larger asymmetry to gain left-right
tunneling as much as possible. (iii) Third, besides the left-right channels
of tunneling there are also same-side tunneling, a larger polaron asymmetry
will enhance the same-side overlap (regions 3,4,5 in Fig.~\ref{fig1}(d))
between the antipolarons and the polarons. These competing factors together
lead to the first reversal asymmetry imbalance.

When the coupling is reduced further one sees another reversal of the
asymmetry imbalance around $g=0.71g_{c}$ in Fig.~\ref{Fig-GS-Vars}(a) (marked by
the other dot). In fact, in such a small-$g$ regime, all the polarons and
antipolarons are very close and nearly in full overlapping. In such a
situation, the competition mainly lies between the left-right tunneling (ii)
and the same-side tunneling (iii), which depends on the difference of the
wave-packet overlap region 5 and region 6. Actually, with the nearly full
overlapping, the variation of region 5 is coming from the peak part which is
less sensitive to the asymmetry changes than the variation of region 6 from
the faster-varying wave-packet sides. A larger $\delta _{\beta }$ can get a
relatively easier increase of tunneling, which leads to the slightly larger
magnitude of $\delta _{\beta }$ again.

\subsection{Unnormal displacement renormalization in attractive polarons and
antipolarons  }

Fig.~\ref{Fig-GS-Vars}(d) gives the evolution of the displacement renormalization
factors $\zeta _{\alpha }$ and $\zeta _{\beta }$. A general tendency is that
$\zeta _{\alpha }$ and $\zeta _{\beta }$ are small before $g_{c}$ but
approach to be $1$ after $g_{c}$, which reflects the process of wave packet
splitting process for the QPT, in the picture that the polarons and
antipolarons tend to stay around the origin before the transition and start
to follow the displacements of the potentials after the transition. Such a
transition-like process is abrupt in the low-frequency limit and softened at
finite frequencies due to the wave-packet broadening~\cite{Ying2015}.
Normally the values of $\zeta _{\alpha }$ and $\zeta _{\beta }$ should be
both smaller than $1$. However, in Fig.~\ref{Fig-GS-Vars}(d) we see at a large
coupling regime, $\zeta _{\alpha }$ umnormally goes beyond the normal limit
value of $1$.

This unnormal behavior of $\zeta _{\alpha }$ originates from
the strong asymmetric effect. In fact, not only the $\zeta _{\alpha }$ can
directly renormalizes the polaron displacement, but also the polaron
asymmetry can indirectly affect the peak position of the polarons due to the
asymmetry-induced wave-packet deformation. Indeed, the final peak positions
of $\varphi _{\alpha }(x)$ and $\varphi _{\beta }(x)$ are located at
\begin{equation}
x_{i}^{p}=\eta _{i}\zeta _{i}g^{\prime }+\frac{\sqrt{\xi _{i}+4\delta
_{i}^{2}}-\sqrt{\xi _{i}}}{2\delta _{i}\sqrt{\xi _{i}}}  \label{x-peak}
\end{equation}%
where $i=\alpha ,\beta $, $\eta _{\alpha }=-1$ and $\eta _{\beta }=1$. The
first term of Eq.~\eqref{x-peak} is the original renormalized displacement,
while second term stems from the polaron asymmetry. We can introduce an
effective displacement renormalization $\zeta _{i}^{p}$ to rewrite the final
peak position as
\begin{equation}
x_{i}^{p}\equiv \eta _{i}\zeta _{i}^{p}g^{\prime }=\eta _{i}\left( \zeta
_{i}+\zeta _{i}^{\delta }\right) g^{\prime },
\end{equation}
with an asymmetry-induced effective renormalization factor
\begin{equation}
\zeta _{i}^{\delta }=\eta _{i}(\sqrt{\xi _{i}+4\delta _{i}^{2}}-\sqrt{\xi
_{i}})/(2\delta _{i}\sqrt{\xi _{i}}g^{\prime }).  \label{Zata-Asymm}
\end{equation}
Note that $\delta _{i}$ has a sign opposite to $\eta _{i}$ and the asymmetry
contributes to discount the displacement. In the large-coupling regime the
polarons have a weight dominant over the antipolaron due to the enlarged
potential difference, in this situation the polarons tend to stay nearly at
the potential bottom, with a weak displacement renormalization. However, the
strong asymmetry creates a relatively large displacement discount which
excessively takes the role of the displacement renormalization. As a result,
the original renormalization $\zeta _{\alpha }$ has to oppositely expand to
partially cancel the overflow of the strong-asymmetry-induced displacement
discount, which leads to a $\zeta _{\alpha }$ larger than $1$. Although this
unnormal $\zeta _{\alpha }$ does not represent the final peak position, it
is a sign that reflects a strong asymmetry effect relatively to the original
displacement renormalization.

\begin{figure}[t]
	\centering
	\includegraphics[width=0.9\columnwidth]{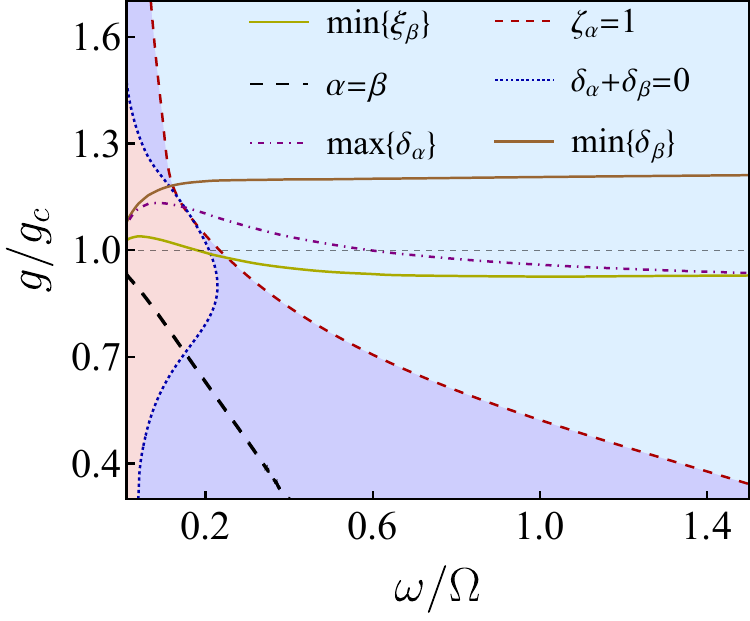}
	\caption{{\it Ground-state phase diagram in the $\omega$-$g$ plane.} The dark-yellow (light gray) solid line denotes the transition around $g_c$ \eqref{Eq-gC} as reflected by the maximum of $\xi_\beta$. The brown (dark gray) solid line around $g=1.2g_c$ (the dot-dashed line) labeled by Min${\delta _\beta}$ (Max${\delta _ \alpha}$) is the maximum asymmetry point of the antipolaron (polaron). The dotted boundary marks the reversal of asymmetry imbalance, with $\delta_\alpha +\delta_\beta >0$ inside the boundary and $\delta_\alpha +\delta_\beta >0$ outside. The dashed line separates the $\zeta_\alpha>0$ region above the line and the $\zeta_\alpha<0$ region below the line. The long-dashed slash is the boundary $\alpha =\beta$ for weight reversal with $\alpha >\beta$ above the boundary and $\alpha <\beta$ below.
 }
	\label{fig4}
\end{figure}

\subsection{Ground-state phase diagram in asymmetric polaron picture}

The ground-state phase diagram is presented in Fig.~\ref{fig4} in the $\omega $-$g$ plane.
The horizontal straight dashed line marks the transition coupling scale $g_{c}$ [Eq.\eqref{Eq-gC}]) at finite frequencies~\cite{Ying2015}, while the
dark-yellow (light gray) solid line around $g=g_{c}$ represents the minimum point of antipolaron frequency
renormalization $\xi _{\beta }$ which reflects the transition-like behavior. The
long-dashed slash denotes the boundary $\alpha =\beta $ for the polaron
weight reversal. Besides these transitions which are the same as in the symmetric polaron picture,
some additional boundaries emerge in the asymmetric polaron picture:
The brown (dark gray) solid line around $g=1.2g_c$ is the minimum point of asymmetry factor $\delta _\beta$ of the antipolarons,
while the dot-dashed line labels the maximum point of $\delta _ \alpha$ of the polarons.
The dotted curve marks the boundary for the asymmetry imbalance reversal, which occurs at low frequencies as the flat potential at
low frequencies fulfils the condition of less steep potential (i) in Sec.~\ref{Sect-asym-reversal}.
The short-dashed curve is the boundary $\zeta _{\alpha }=1$ for the unnormal displacement renormalization. The phase diagram provides an
overall view for the subtle energy competitions in the ground state and abundant physics
explored by the asymmetric polaron picture.

\begin{figure}[t]
	\centering
	\includegraphics[width=1.0\columnwidth]{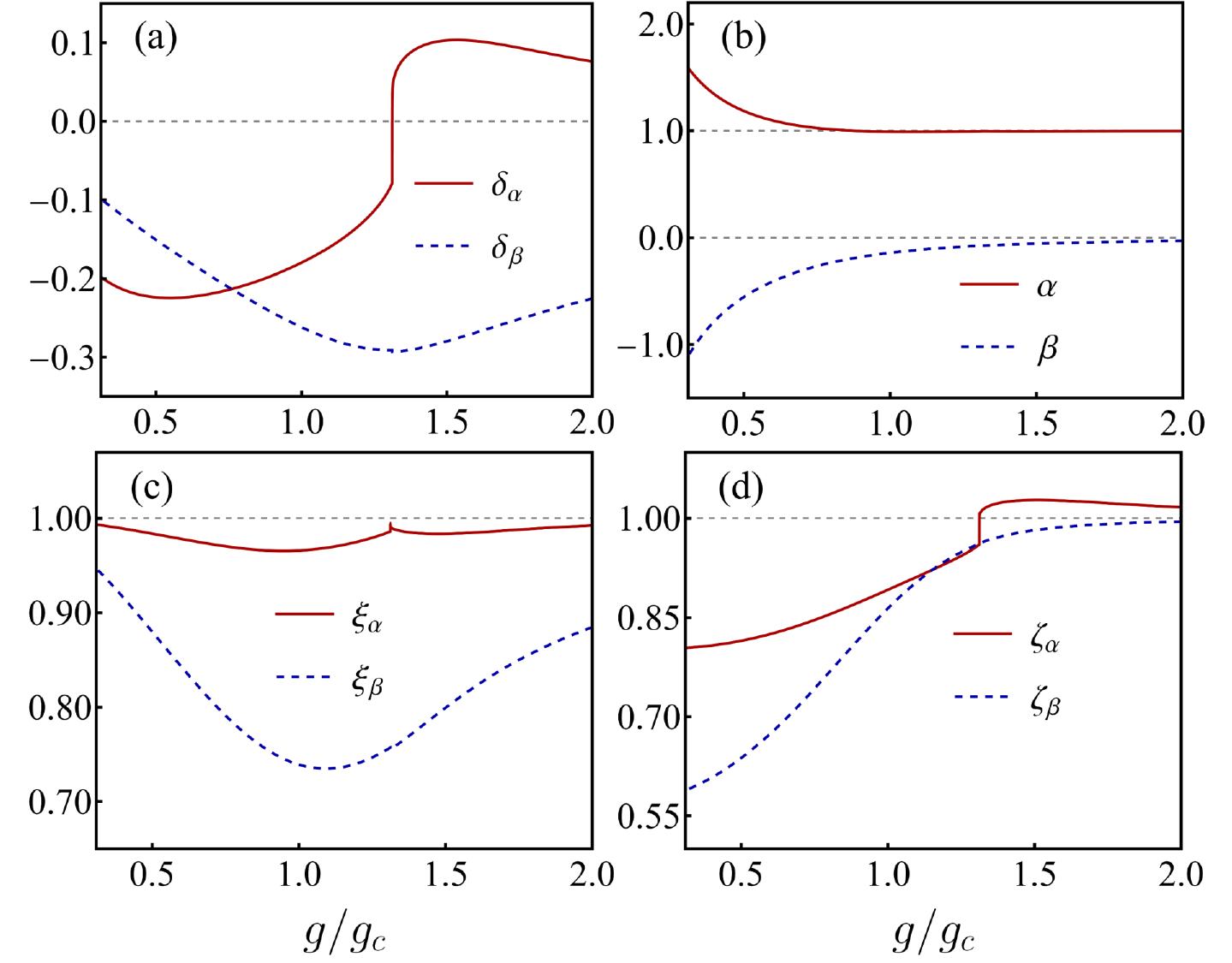}
	\caption{{\it Variational parameters and transitions in the first excited state.} Evolution of the variational parameters for (a) asymmetry factors $\delta_\alpha$ (solid) and $\delta_\beta$ (dashed), (b) weights of polaron ($\alpha$, solid) and antipolaron ($\beta$, dashed), (c) renormalization factors of polaron ($\xi_\alpha$, solid) and antipolaron ($\xi_\beta$, dashed), (d) displacement renormalization factors of polaron ($\zeta_\alpha$, solid) and antipolaron ($\zeta_\beta$, dashed). Here $\omega=0.5\Omega$.
}
	\label{fig5}
\end{figure}

\begin{figure}[t]
	\centering
	\includegraphics[width=1.0\columnwidth]{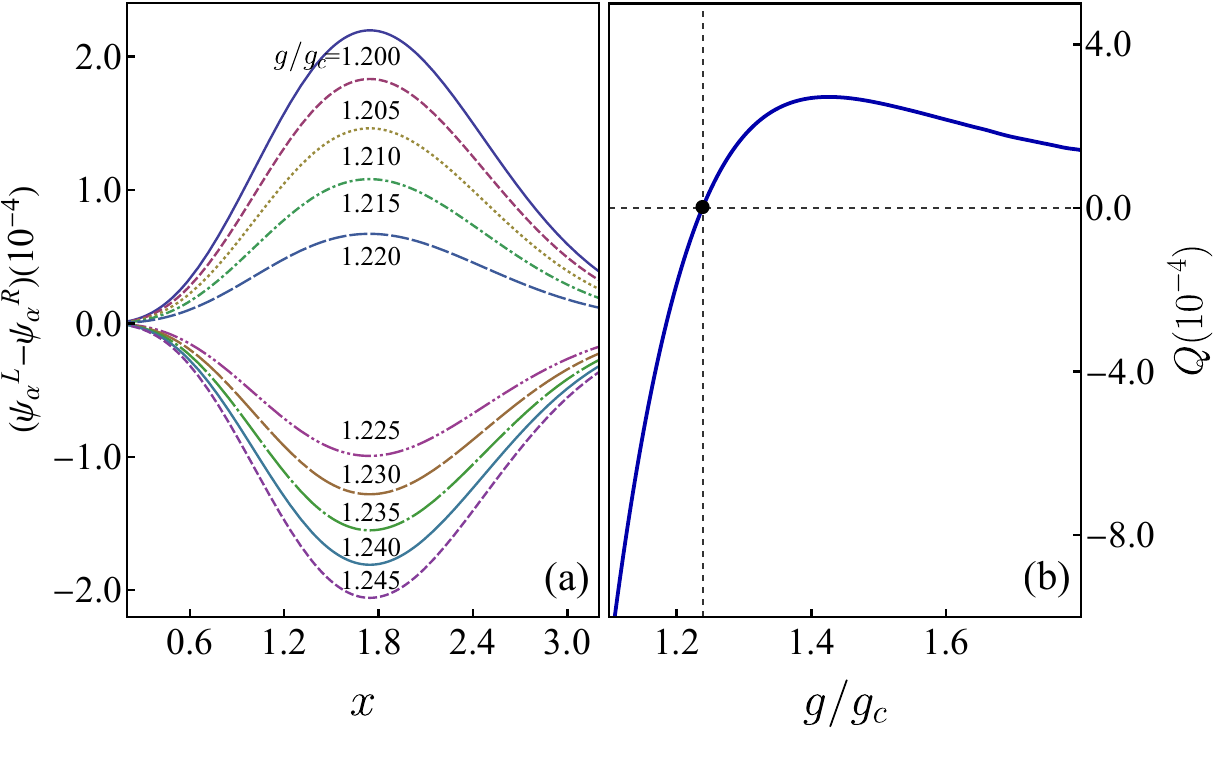}
	\caption{{\it Transition of asymmetry sign reversal confirmed by the ED results for the first excited state.} (a) Asymmetry of polaron wave packet by difference between its left side ($\psi_\alpha^L$) and right side ($\psi_\alpha^R$) as a function of the distance $x$ from the peak, at couplings around the point of asymmetry sign reversal. The numbers give the coupling strength $g/g_c$. Here $\psi_\alpha$ is extracted by polaron fitting of the ED wave function. (b) The sign reversal of the asymmetry quantity $Q$ (Eq.\eqref{def-Q}) from the ED wave function. Here $\omega=0.5\Omega$.}
	\label{fig6}
\end{figure}

\section{The first excited state in asymmetric polaron picture\label{Sect-Excited-State}}

Now we apply the variational method of asymmetric polaron picture to study
the first excited state. We will see that the first excited state manifests
some transitions and physics differently from the ground state.

\subsection{Opposite weight signs of polarons and antipolarons}

Fig.~\ref{fig5}(b) shows the weights of the polarons and antipolarons. At a
first glance one can immediately see the first difference from the ground
state that the weights of the polarons and the antipolarons here have
opposite signs, as $\beta $ turns to be negative. This comes form the parity difference, as the ground state
has a negative parity $P=-1$ while the first excited state possesses a
positive parity $P=+1$. Although the energy is higher than the ground state,
the first excited state is the lowest state within all the states in
positive parity. Under the positive parity, the opposite signs of the
polarons and antipolarons guarantee a negative tunneling energy in the
same-side channels $\Omega _{\alpha \beta }=\frac{\Omega }{2}\alpha \beta
P\langle \varphi _{\alpha }(x)|\overline{\varphi }_{\beta }(x)\rangle $ and
$\Omega _{\alpha \beta }=\frac{\Omega }{2}\alpha \beta P\langle \overline{%
\varphi }_{\beta }(x)|\varphi _{\alpha }(x)\rangle ,$ despite that the
left-right channels, $\Omega _{\alpha \alpha }=\frac{\Omega }{2}\alpha
^{2}P\langle \varphi _{\alpha }(x)|\overline{\varphi }_{\alpha }(x)\rangle $
and $\Omega _{\beta \beta }=\frac{\Omega }{2}\beta ^{2}P\langle \overline{%
\varphi }_{\beta }(x)|\varphi _{\beta }(x)\rangle $, have a positive
tunneling energy which however has a smaller contribution than the
same-side channels due to the smaller wave-packet overlaps $\langle \varphi
_{\alpha }(x)|\overline{\varphi }_{\alpha }(x)\rangle $ and $\langle
\overline{\varphi }_{\beta }(x)|\varphi _{\beta }(x)\rangle $.

\subsection{Weight magnitudes both increasing for polarons and antipolarons
in reduced coupling}

With a further observation on the weight evolution one may realize that, when the
coupling strength is reduced, the weight magnitudes of the polarons and the
antipolarons, $\left\vert \alpha \right\vert $ and $\left\vert \beta
\right\vert $, are both increasing (subject to the normalization). In
contrast, in the ground state the polaron weight is decreasing while
antipolaron weight is increasing, so that $\alpha $ and $\beta $ have a
similar amplitude at small values of $g$. This special weight magnitude
behavior comes from the parity difference. as the wave-function
normalization requires $\left\langle \psi |\psi \right\rangle=
\alpha ^{2}  \langle \varphi _{\alpha }(x)|\varphi _{\alpha }(x)\rangle
+\beta ^{2}  \langle \varphi _{\beta }(x)|\varphi _{\beta }(x)\rangle
-P\left\vert \alpha \beta \right\vert \langle \varphi _{\alpha }(x)|\varphi _{\beta }(x)\rangle =1$.
We know that a smaller coupling leads to a larger overlap
$\langle \varphi _{\alpha }(x)|\varphi _{\beta }(x)\rangle $ in a closer polaron-antipolaron
distance and an increasing antipolaron weight $\left\vert \beta \right\vert $
in a lower potential, while the same-side overlaps
$\langle \varphi _{\alpha }(x)|\varphi _{\alpha }(x)\rangle$ and
$\langle \varphi _{\beta }(x)|\varphi _{\beta }(x)\rangle$ are much
less affected by the left-right distance and the left-right potential difference. In such a situation, the ground state with $P=-1$ needs
a decreasing polaron weight $\left\vert \beta \right\vert $ to fulfill the
normalization condition, while this trend is reversed for the first excited
state with $P=+1$ due to the third term of $\left\langle \psi |\psi \right\rangle$ now becomes negatively increasing.
Therefore, we have increasing weight magnitudes for both polarons and antipolarons.

\subsection{Transition of polaron attraction and repulsion}\label{Sect-delta-reversal}

In the first excited state the transition around $g_{c}$ reflected by
maximum frequency renormalization (minima of $\xi _{\alpha }$ and $\xi
_{\beta }$) is similar to the ground state, as illustrated by Fig.~\ref{fig5}(c).
A transition different from the ground state arises as the asymmetry
factor $\delta _{\alpha }$ changes sign at a certain point when the coupling
is reduced, as shown in Fig.~\ref{fig5}(a). Note that the sign reversal in Sec.~\ref{Sect-asym-reversal}
for the ground state is for the imbalance between
two asymmetry factors $\delta _{\alpha }$ and $\delta _{\beta }$, while here
the sign reversal occurs in one asymmetry factor $\delta _{\alpha }$. The $%
\delta _{\alpha }$ sign shifting from positive to negative means that some
distribution probability of the polarons are moved from the sides close to
the origin to the opposite sides far away from the origin. This effect
appears to be a change from attraction to repulsion. Indeed, the left and
right polarons have a trend to avoid overlapping due to the positive
tunneling energy of $\Omega _{\alpha \alpha }$ which is increasing due to
the enlarged left-right polaron overlap $\langle \varphi _{\alpha }(x)|%
\overline{\varphi }_{\alpha }(x)\rangle $ in closer polaron distances when
the coupling is reduced. The sign reversal of $\delta _{\alpha }$ reduces $%
\langle \varphi _{\alpha }(x)|\overline{\varphi }_{\alpha }(x)\rangle $ to a
largest degree. This sign transition does not happen to $\delta _{\beta }$
due to the potential-induced asymmetry is stronger in antipolarons.

\begin{figure}[t]
	\centering
    \includegraphics[width=0.7\columnwidth]{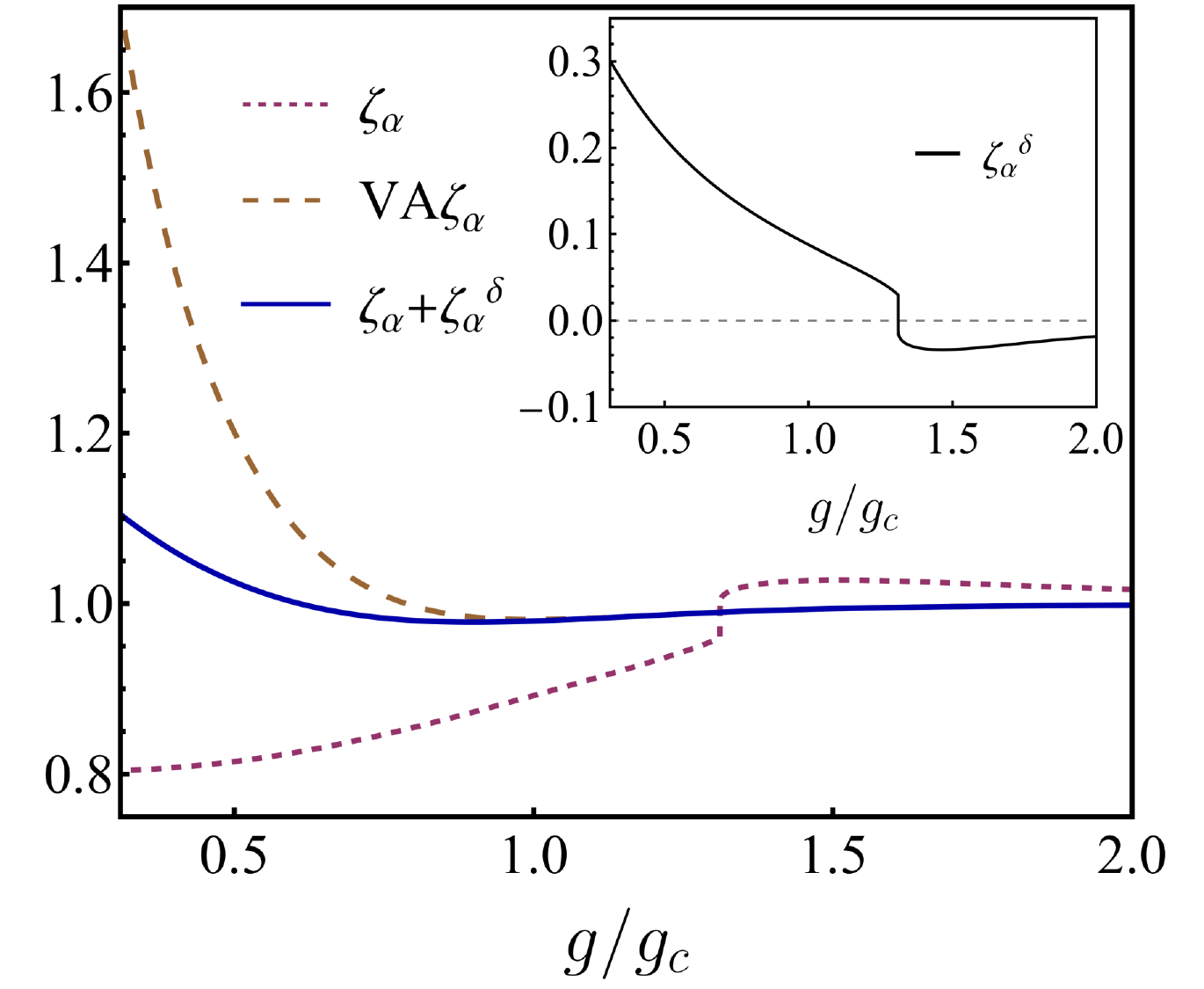}
	\caption{{\it Transition of polaron attraction and repulsion in
the first excited state.}  The displacement factors from variational parameter $\zeta _\alpha$ (dashed), from asymmetry $\zeta_\alpha^d$ (Inset), and from the total one $\zeta_\alpha+\zeta_\alpha^\delta$ (solid),  in the case of asymmetry reversal in Fig. \ref{fig6}. }
	\label{fig9}
\end{figure}

The sign reversal of $\delta _{\alpha }$ in the above is observed from the
evolution of $\delta _{\alpha }$ which is a variational parameter in energy
minimization. This asymmetry transition might be further confirmed by
results from exact diagonalization (ED). First, we use the variational wave
function to fit the ED wave function, from which we can extract the
asymmetry of the wavepacket, as indicated in Fig.~\ref{fig6}(a) by the
left-right difference around the main-peak position corresponding to the
polaron. We see that indeed the left-right difference reverses the sign
around the same transition point in Fig.~\ref{fig5}(a).

To further verify this
asymmetry reversal, we introduce an asymmetry quantity
\begin{equation}
Q=\int_{-\infty }^{\infty }\psi _{{\rm ED}}^{\ast
}(x)\ {\rm sign}(x-x_{m})e^{-(x-x_{m})^{2}/2}\psi _{{\rm ED}}(x)dx  \label{def-Q}
\end{equation}%
where ${\rm sign}(x)$ is the sign of $x$ and $x_{m}$ is the main-peak position of
the ED wave function $\psi _{{\rm ED}}(x)$. Here in the integrand the factor
$e^{-(x-x_{m})^{2}/2}$ filters the affection from the part far away from the
peak, which corresponds to the antipolaron in the variational method, and
amplifies contribution from the part around the peak, which corresponds to
the polaron. The asymmetry is then captured by adding ${\rm sign}(x-x_{m})$ as a
coefficient of $\psi _{{\rm ED}}(x)$: $Q$ would be canceled if $\psi _{{\rm %
ED}}(x)$ is symmetric around the the peak, otherwise the sign of $Q$
reflects different direction of asymmetry with higher density on the right
side of the peak for $Q>0$ and higher density on the left side for $Q<0$. It
turns out that there is really a sign reversal in $Q$ as shown by Fig.~\ref%
{fig6}(b), which occurs around the same point of the afore-mentioned
asymmetry transition in $\delta _{\alpha }$ sign reversal.

\begin{figure}[t]
	\centering
	\includegraphics[width=1.0\columnwidth]{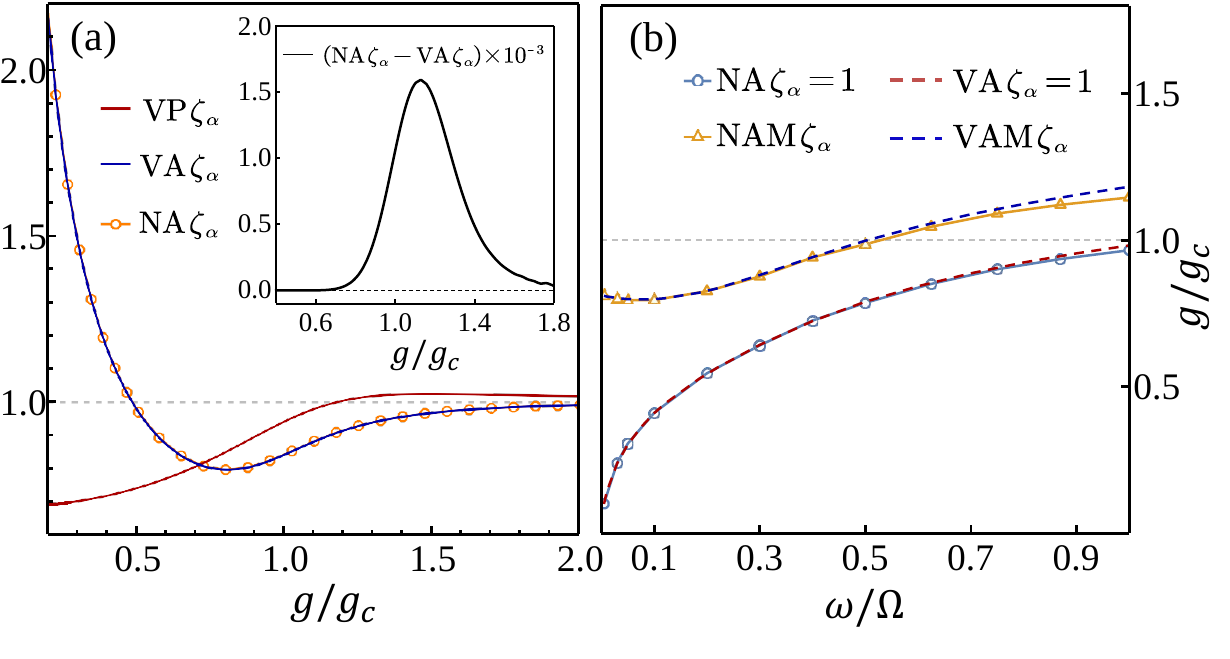}
	\caption{{\it Polaron-density-cancellation-induced repulsion in
the first excited state.} (a) Displacement factors of the main peak position by the numerically accurate wave function (NA $\zeta _\alpha$, circles) from the ED and by variational polaron fitting of the accurate ED wave function (VA $\zeta _\alpha$, solid line), at $\omega=0.15\Omega$. The inset shows their discrepancy. (b) Frequency dependence of the increasing/decreasing turning point or minimum point (NAM $\zeta _\alpha$ and VAM $\zeta _\alpha$,  triangles and dashed line) and the point going beyond $1$ (NA $\zeta _\alpha=1$ and VA $\zeta _\alpha=1$, circles and long-dashed line) in (a).}
	\label{fig7-repulsion}
\end{figure}

\begin{figure}[ht]
	\centering
    \includegraphics[width=0.9\columnwidth]{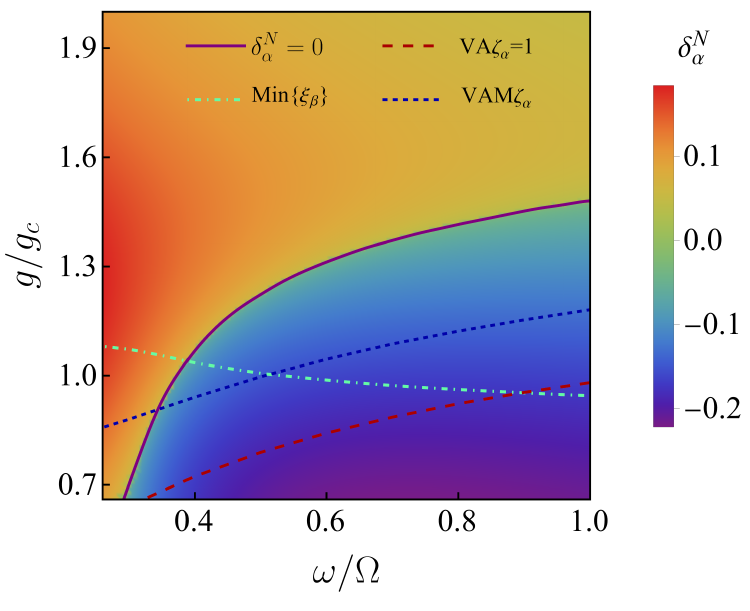}
	\caption{{\it Phase diagram of the first excited state in the $\omega$-$g$ plane.} The dot-dashed line represents the transition around $g_c$, as reflected by the minimum of $\xi _\beta$. The density plot is for $\delta _\alpha$ by polaron fitting of the numerical ED wave function (denoted by $\delta _\alpha^N$). The solid line marks the additional transition boundary for the sign reversal of $\delta _\alpha$. The dashed line and dotted line are the minimum point and the point going beyond $1$ of main-peak $\zeta _\alpha$ in Fig.\ref{fig7-repulsion}(a).}
	\label{fig8-excitedS-phase-diagam}
\end{figure}

\subsection{Polaron-density-cancellation-induced repulsion in positive parity}

The polaron repulsive effect in Sec.\ref{Sect-delta-reversal} is coming from the
positive tunneling energy of $\Omega _{\alpha \alpha }$ in positive parity
of the first excited state. The positive parity also leads to a ``repulsion"
effect from density cancellation. In fact the opposite signs of the weights $\alpha $
and $\beta $ results in a wave-function cancellation in the meeting part of
the polaron and the antipolaron, which reduces the density in the valley
between the main polaron peak and the secondary antipolaron peak. Thus the
polaron peak and the antipolaron peak are pushed farther away from each
other, especially when they are close in the small couplings. This effect
also appears to be repulsive-like.

Fig.~\ref{fig7-repulsion}(a) demonstrates
the evolution of the final position of the main polaron peak by the line
with circles, the effective displacement renormalization factor $\zeta
_{\alpha }$ for which is turning from decreasing to increasing in coupling
reduction and even becomes diverging at small couplings despite that the
variatonal $\zeta _{\alpha }$ (solid line without circles) is always
decreasing and smaller than $1$. Fig.~\ref{fig7-repulsion}(b) shows the
frequency dependence of the increasing/decreasing turning point (triangles)
of the final peak position and the point beyond $1$ (circles). This
``repulsion" effect is in a sharp contrast to the ground state where the
polarons and antipolarons are attractive-like and their final peak position never goes
beyond $1$.

\subsection{Phase diagram of the first excited state}

Fig.~\ref{fig8-excitedS-phase-diagam} shows the phase diagram in the $\omega
$-$g$ plane for the first excited state from the above discussions. The
dot-dashed line represents the minimum point of $\xi _{\beta }$, which
denotes the transition around $g_{c}$ similar to the ground state. The
density plot in Fig.~\ref{fig8-excitedS-phase-diagam} provides the map of
$\delta _{\alpha }$ numerically extracted (denoted by $\delta _\alpha^N$) from the fitting of the ED wave function
illustrated in Fig.~\ref{fig6}(a), with the sign reversal boundary (solid
line) separating the positive (red, above the solid line) and negative
(blue, below the solid line) regions. The transition of $\delta _{\alpha }$
sign reversal is different from $g_{c}$, occurring below $g_{c}$ at low
frequencies and above $g_{c}$ at high frequencies. The dashed line and the
dotted line locate the increasing/decreasing turning point of the final main
peak position and the point beyond $1$ for the final peak displacement
factor $\zeta _{\alpha }$, which reflects the degree of the effective
polaron repulsion induced by the density cancellation in positive parity.

\section{Application in quantum Fisher information}
\label{Sect-Fq-gc}

\subsection{Quantum Fisher information (QFI) for quantum metrology}
\label{Section-QFI}

As established by the Cram\'{e}r-Rao theorem~\cite{Cramer-Rao-bound} the
measurement precision of experimental estimation on a parameter $\lambda $
is bounded by $F_{Q}^{1/2}$ in quantum metrology. Here $F_{Q}$ is the
quantum Fisher information (QFI) defined as \cite{Cramer-Rao-bound,Taddei2013FisherInfo,RamsPRX2018}
\begin{equation}
F_{Q}\left( \lambda \right) =4\left[ \langle \psi ^{\prime }\left( \lambda
\right) |\psi ^{\prime }\left( \lambda \right) \rangle -\left\vert \langle
\psi ^{\prime }\left( \lambda \right) |\psi \left( \lambda \right) \rangle
\right\vert ^{2}\right]   \label{Fq}
\end{equation}%
for a pure state $|\psi (\lambda )\rangle $ and $^{\prime }$ denotes the
derivative with respect to the parameter $\lambda $. A larger QFI would mean
a higher precision accessible in measurements. For a real wave function $%
\psi (\lambda )$, as is usually the case in non-degenerate states of a real
Hamiltonian, the second term in Eq.~\eqref{Fq} vanishes so that the QFI can
be simplified to be~\cite{Ying-gC-by-QFI-2024}
\begin{equation}
F_{Q}=4\langle \psi ^{\prime }\left( \lambda \right) |\psi ^{\prime }\left(
\lambda \right) \rangle ,  \label{Fq-Simplified}
\end{equation}
which also applies for the ground state of the QRM (\ref{eq:rabi})
considered in the present work.

The appearance of QFI peak in QPTs can be employed for critical quantum
metrology~\cite{Garbe2020,Montenegro2021-Metrology,Chu2021-Metrology,Garbe2021-Metrology,Ilias2022-Metrology, Ying2022-Metrology,YangZheng2023SciChina,Gietka2023PRL-Squeezing,Hotter2024-Metrology,Alushi2024PRL,Mukhopadhyay2024PRL,Mihailescuy2024,
Ying-Topo-JC-nonHermitian-Fisher,*Ying-Topo-JC-nonHermitian-Fisher-Cover,Ying-g2hz-QFI-2024,*Ying-g2hz-QFI-2024-Cover,
Ying-g1g2hz-QFI-2025,Ying2025g2A4,*Ying2025g2A4-Cover,Ying-g2Stark-QFI-2025,Gietka2025PRL100802,Mihailescu2025CQMtutorial,QiuYi2025gA2}.
Conversely, such QFI peaks have been applied to identify the QPT in
light-matter interactions~\cite{Ying-gC-by-QFI-2024}. In fact, the QFI is
equivalent to the susceptibility of the fidelity. In an infinitesimal
parameter variation $\delta \lambda $ the fidelity $F$ can be expanded as
\begin{equation}
F=\left\vert \langle \psi \left( \lambda \right) |\psi \left( \lambda
+\delta \lambda \right) \rangle \right\vert =1-\frac{\delta \lambda ^{2}}{2}
\chi _{F},
\end{equation}%
so that the QFI corresponds to the susceptibility of the fidelity by
$\chi_{F}=F_{Q}/4$~\cite{Gu-FidelityQPT-2010,You-FidelityQPT-2007,You-FidelityQPT-2015,Ying-g1g2hz-QFI-2025}.
The appearance of the QFI peak signals a QFT in fidelity
theory~\cite{Zhou-FidelityQPT-2008,Gu-FidelityQPT-2010,You-FidelityQPT-2007,You-FidelityQPT-2015,Zanardi-FidelityQPT-2006}.

In the following we shall apply the asymmetric polaron picture to analyze
the QFI of the QRM.

\begin{figure}[t]
	\centering
    \includegraphics[width=1\columnwidth]{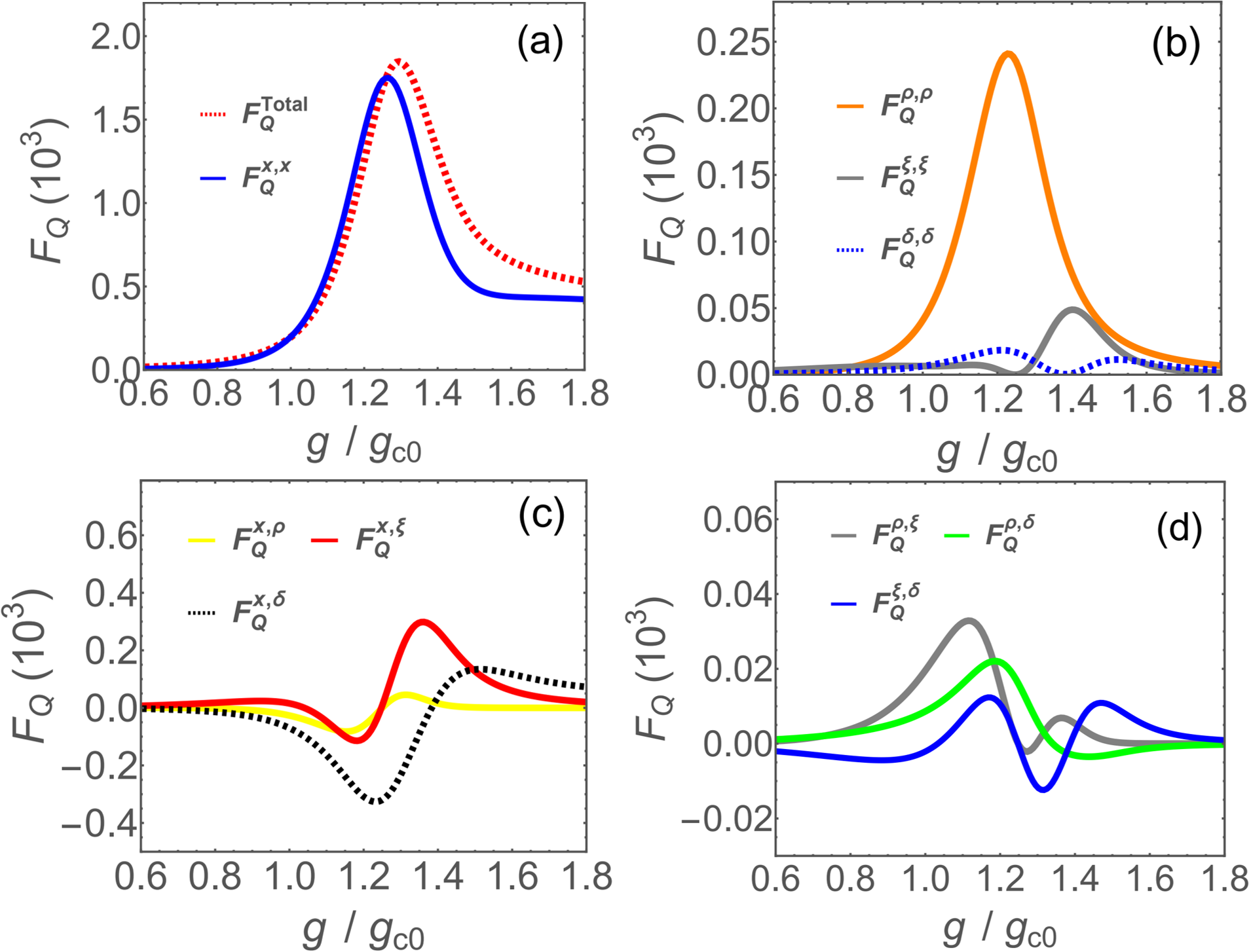}
	\caption{{\it Decomposed quantum Fisher information (QFI) of the ground state and contribution of polaron asymmetry.}
(a) The total QFI $F_Q^{\rm Total}$ (red dotted line) and the displacement part $F_Q^{ x,x}$ (blue solid line).
(b) Decomposed QFI from variations of polaron weight [$F_Q^{\rho,\rho}$, orange (light gray) solid line],
frequency renormalization [$F_Q^{\xi,\xi}$, gray solid line] and polaron asymmetry [$F_Q^{\delta,\delta}$, blue dotted line].
(c) Interplay parts of QFI between displacement and polaron weight [$F_Q^{x,\rho}$, red (dark gray) solid line],
frequency renormalization [$F_Q^{x,\xi}$, green (light gray) solid line] and polaron asymmetry [$F_Q^{x,\delta}$, black dotted line].
(d) Interplay parts of QFI between polaron weight and frequency renormalization [$F_Q^{\rho,\xi}$, gray solid line],
polaron weight and asymmetry [$F_Q^{\rho,\delta}$, green (light gray) solid line] and frequency renormalization and asymmetry [$F_Q^{\xi,\delta}$, blue (dark gray) solid line]. Here $\omega/\Omega=0.1$ and $g_{c0}=\sqrt{\omega\Omega}/2$ is the critical coupling in low-frequency limit.
}
	\label{Fig-Fq-decomposed}
\end{figure}
\begin{figure}[t]
	\centering
    \includegraphics[width=1\columnwidth]{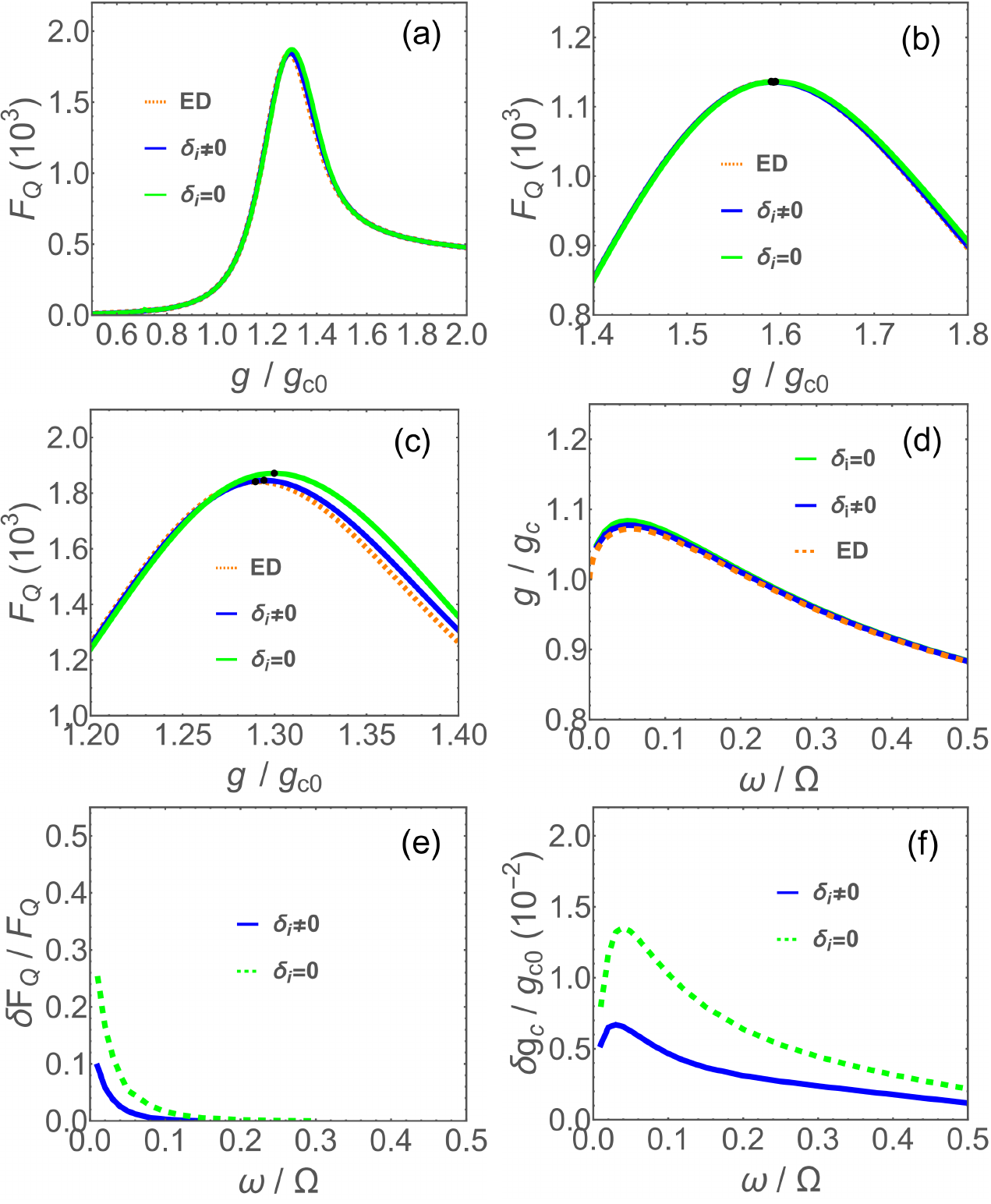}
	\caption{{\it Improvements on QFI and critical coupling from polaron asymmetry.}
(a) The QFI in symmetric ($\delta _i=0$) and asymmetric ($\delta _i\neq 0$) polaron pictures compared to the exact diagonalization (ED) at $\omega/\Omega=0.1$.
(b) The QFI around the peak position at $\omega/\Omega=0.3$.
(c) Close-up view of the QFI around the peak position at $\omega/\Omega=0.1$.
(d) The critical couplings versus frequency extracted by the QFI peak position for the symmetric and asymmetric polaron pictures compared to the ED result in Eq.~\eqref{Eq-gC-by-Fq}.
(e) Discrepancy of the QFI peak values from the ED.
(f) Discrepancy of the critical couplings from the ED.
}
	\label{Fig-Fq-gc}
\end{figure}

\subsection{Decomposed QFI and contribution of polaron asymmetry}

In the asymmetric polaron picture, we have variation resources from
displacement ($x$), frequency renormalization ($\xi $), polaron asymmetry ($\delta $)
and polaron weight ($\rho $). The ground state of the the QRM is
non-degenerate so that the QFI takes the simplified form (\ref{Fq-Simplified}).
Then, with the four variation resources, the QFI can be decomposed into
different parts:
\begin{eqnarray}
F_{Q} &=&F_{Q}^{x,x}+F_{Q}^{\xi ,\xi }+F_{Q}^{\delta ,\delta }+F_{Q}^{\rho
,\rho }  \nonumber \\
&&+F_{Q}^{x,\xi }+F_{Q}^{x,\rho }+F_{Q}^{x,\delta }+F_{Q}^{\rho ,\xi
}+F_{Q}^{\rho ,\delta }+F_{Q}^{\xi ,\delta },
\end{eqnarray}
where the first four terms are purely intra-resource contributions, while
the others terms come from inter-resource interplays. We leave the
definitions of these term in Appendix \ref{Appendix-Fq-decomposed}.

We illustrate the different contributions of QFI in Fig.\ref{Fig-Fq-decomposed}.
We see in Fig.\ref{Fig-Fq-decomposed}(a) that the pure displacement part
$F_{Q}^{x,x}$ (blue solid line) contribute a main portion in the total QFI
$F_{Q}^{{\rm Total}}$ (dotted red line), while the intra-resource parts
($F_{Q}^{\xi ,\xi }$, $F_{Q}^{\delta ,\delta }$, $F_{Q}^{\rho ,\rho }$) make
relatively smaller contributions as shown in Fig.~\ref{Fig-Fq-decomposed}(b).
Still, despite that the inter-resource terms $F_{Q}^{\rho ,\xi }$,
$F_{Q}^{\rho ,\delta }$, $F_{Q}^{\xi ,\delta }$ are small as in Fig.~\ref{Fig-Fq-decomposed}(d),
we find in Fig.~\ref{Fig-Fq-decomposed}(c) that the
polaron asymmetry (black dotted line) gives a considerable contribution via
the interplay with the displacement.

\subsection{Improvements on QFI and critical couplings}

We finally present the results in symmetric and asymmetric polaron pictures
in comparing the QFI and critical couplings with the numerically accurate
results in ED. Indeed, as illustrated in Fig.\ref{Fig-Fq-decomposed}(a),
both the symmetric [$\delta _{i}=0$, green (light gray) solid line] and
asymmetric [$\delta _{i}\neq 0$, green (dark gray) solid line] polaron
pictures can capture the variations of the QFI extracted in ED (orange
dotted line). At finite frequencies both the symmetric and asymmetric
polaron pictures yield fairly accurate results illustrated by the frequency
example $\omega /\Omega =0.3$ in Fig.~\ref{Fig-Fq-decomposed}(b). However,
discrepancy from the ED results arises in lower frequencies as $\omega
/\Omega =0.1$ in Fig.~\ref{Fig-Fq-decomposed}(c) in the close-up plot. Still,
here we see that the asymmetric polaron picture leads to smaller discrepancy
than the symmetric polaron picture, with both the peak values and positions
of the QFI closer to that of ED as marked by the dots.

Figure \ref{Fig-Fq-decomposed}(d) gives the frequency dependence of the critical
couplings extracted from the peak position of the QFI. Indeed, the critical
couplings at finite frequencies are not any more located at the
low-frequency one $g_{c0}$ as illustrated in Fig.~\ref{Fig-Fq-decomposed}(a), while $g_{c}$ in Eq.~\eqref{Eq-gC}, taken as the
unit in the Fig.~\ref{Fig-Fq-decomposed}(d), captures better the right scale.
Still, the accurate one is given by $g_{c}^{{\rm F}}$ (dashed line) in Eqs.~\eqref{gc2} and \eqref{Eq-gC-by-Fq}. We see in the entire
frequency regime that asymmetric polaron picture (blue solid line)
considerably reduces the errors of the QFI peak values in Fig.~\ref{Fig-Fq-decomposed}(e) and critical-like couplings
in Fig.~\ref{Fig-Fq-decomposed}(f). There is still some remnant errors, especially for the QFI peak values
in low frequency limit, as we have only considered the lowest order of
polaron asymmetry as in Eq.~\eqref{A-factor}. The errors can be further
reduced by including higher order approximations. Nevertheless, these
results are enough to demonstrate that the effect of polaron asymmetry is
innegligible and richer physics emerges in the presence of the polaron asymmetry.

\section{Conclusions\label{Sect-Conclusions}}

We have proposed the variational method in asymmetric polaron picture for
the QRM. The method is capable of capturing the physics of
polaron asymmetry arising from the tunneling energy and the potential
gradient in the subtle energy competitions, which is missing in the previous
symmetric polaron picture. We have applied the method to study the ground
state and first excited state. The results show that the asymmetric polaron
picture not only yields accuracy improvements for the physical quantities
but also enables us to extract several underlying transitions unnoticed in
the symmetric polaron picture.

For the ground state we have found that the asymmetry effect for the
polarons and antipolarons are quite considerable. We have seen the opposite
asymmetry signs for the polarons and antipolarons, which indicates they are effectively
attracting each other. Two transitions of asymmetry imbalance reversal
appear at low frequencies, which reflects the delicate competition between
the tunneling-induced asymmetry and the potential-gradient-induced potential
asymmetry. We also notice that the polaron displacement is not only decided
by the direct displacement renormalization but also much affected by the
polaron asymmetry. We figure out a boundary of unnormal displacement
renormalization beyond $1$ existing for all frequencies, which is a sign of
strong polaron asymmetry.

Despite that variational methods usually apply for the ground state, we have
also utilized the asymmetric polaron picture to analyze the first excited
state with good outcomes. Besides the maintained accuracy, we have unveiled
some distinguished features and additional transitions different from the
ground state. Indeed, the weights of the polarons and antipolarons have
opposite signs and the magnitudes are both increasing in coupling reduction,
in contrast to the same signs and decreasing polaron weight magnitude in the
ground state. Apart from the main transition around $g_{c}$ the same as the
ground state, we have also found a transition of sign reversal for the
polaron asymmetry factor, which indicates an attraction/repulsion shift for
the polarons. Moreover, the positive parity also leads to a density
cancellation between polarons and antipolarons, which results in a more
explicit repulsion effect with a diverging displacement renormalization at
small couplings for the final polaron peak.

We have finally applied the asymmetric polaron picture in quantum Fisher information analysis
and critical coupling extraction, with improvements over the symmetric polaron picture in both these two properties.
The improvements indicate that the polaron asymmetry makes a
considerable contribution to the quantum resource in quantum metrology and plays an unnegligible
role in the QPT.

Although the components of the QRM are finite and the model
has been much studied~\cite{
Braak2011,Solano2011,Boite2020,Liu2021AQT,Diaz2019RevModPhy,Kockum2019NRP,Rabi-Braak,Braak2019Symmetry,
Wolf2012,FelicettiPRL2020,PengJ2021PRL,
Felicetti2018-mixed-TPP-SPP,Felicetti2015-TwoPhotonProcess,Simone2018,Alushi2023PRX,
Irish2014,Irish2017,Irish-class-quan-corresp,
PRX-Xie-Anistropy,Batchelor2015,XieQ-2017JPA,
Hwang2015PRL,Bera2014Polaron,Hwang2016PRL,Ying2015,LiuM2017PRL,Ying-2018-arxiv,Ying-2021-AQT,Ying-gapped-top,Ying-Stark-top,Ying-Spin-Winding,Ying-JCwinding,Grimaudo2022q2QPT,Grimaudo2023-Entropy,
CongLei2017,CongLei2019,Ying2020-nonlinear-bias,ChenQH2012,
e-collpase-Garbe-2017,e-collpase-Duan-2016,Garbe2020,Rico2020,
Garbe2021-Metrology,Chu2021-Metrology,Ilias2022-Metrology,Ying2022-Metrology,
Boite2016-Photon-Blockade,Ridolfo2012-Photon-Blockade,Li2020conical,
Ma2020Nonlinear,
ZhangYY2016,ZhengHang2017,Yan2023-AQT,Zheng2017,Chen-2021-NC,Lu-2018-1,Gao2021,PengJie2019,Liu2015,Ashhab2013, ChenGang2011-GVM,ChenGang2012,FengMang2013,Eckle-2017JPA,Maciejewski-Stark,Xie2019-Stark,Casanova2018npj,HiddenSymMangazeev2021,HiddenSymLi2021,HiddenSymBustos2021,
JC-Larson2021,Stark-Cong2020,Cong2022Peter,Stark-Grimsmo2013,Stark-Grimsmo2014}, our more-extracted results hint that the physics we could learn from
the model might be infinite. The polaron picture is a useful method to analyze the
QPTs in light-matter models~\cite{Liu2021AQT,Ying2015,Ying2020-nonlinear-bias,CongLei2019,Ying-2018-arxiv,Ying-2021-AQT,Ying-gapped-top,
Ying-Stark-top,CongLei2017,CongLei2019,Ying-2018-arxiv,Ying-gC-by-QFI-2024,
Ying-g2hz-QFI-2024,*Ying-g2hz-QFI-2024-Cover,
Ying-g1g2hz-QFI-2025,Ying-g2Stark-QFI-2025}, with high accuracy and transparent
physical picture. The asymmetric polaron picture we develop in this paper
might help to deepen the physics exploration in road of reaching the ideal goal of full
understanding.

As a final remark, although the present work is focusing on the QRM, our analysis can be extended to other light-matter models. Indeed, in more general models with the anisotropic coupling~\cite{PRX-Xie-Anistropy,Forn-Diaz2010,Pietikainen2017,Yimin2018,Wang2019Anisotropy,LiuM2017PRL,Ying-2021-AQT,Ying-gapped-top}
and the Stark coupling~\cite{Eckle-2017JPA,Stark-Grimsmo2013,Stark-Grimsmo2014,Xie2019-Stark,Stark-Cong2020,Maciejewski-Stark,Ying-Stark-top,*Ying-Stark-top-Cover,Ying-g2Stark-QFI-2025,Ying-JCwinding}, the ground state before first topological transition is similar to that of the QRM considered here, while the second phase after the first topological transition is similar to the first excited state of the QRM~\cite{Ying-2021-AQT,Ying-gapped-top,Ying-Stark-top,Ying-Spin-Winding,Ying-JCwinding}. Also, extensions to multiple-polaron case~\cite{CongLei2017,Ying-gapped-top} and the frequency dimension in nonlinear QRM~\cite{CongLei2019,Ying-g2hz-QFI-2024,*Ying-g2hz-QFI-2024-Cover,
Ying-g1g2hz-QFI-2025,Ying2025g2A4,*Ying2025g2A4-Cover,Ying-g2Stark-QFI-2025} are possible. These examples indicate that our asymmetric polaron picture could have broader applications, which may need more future works.

\section*{Acknowledgment}

This work was supported by the National Natural Science Foundation of China
(Grants No. 12474358 and No. 12247101).

\appendix

\section{Decomposed QFI in asymmetric polaron picture}

\label{Appendix-Fq-decomposed}

As addressed in Sec.\ref{Section-QFI}, the QFI provides the upper bound of
measurement precision in quantum metrology. Here in this Appendix we
formulate the QFI in the asymmetric polaron picture developed in the main
text. The wave function cane be decomposed into a linear combination of $%
n_{p}$ number of polarons
\begin{equation}
\psi _{\pm }(x)=\sum\limits_{a=1}^{n_{p}}c_{a}^{\pm }\varphi _{a}^{\pm }(x)
\end{equation}%
where%
\begin{equation}
\varphi _{a}^{\sigma }(x)=(1+\delta _{a}^{\sigma }x_{a}^{\sigma })\left(
\frac{\xi _{a}^{\sigma }}{\pi }\right) ^{1/4}\exp [-\frac{1}{2}\xi
_{a}^{\sigma }(x+x_{a}^{\sigma })^{2}]  \label{polaron-general}
\end{equation}%
represents an asymmetric polaron labelled by $a$ in the spin component $%
\sigma =\pm $. Generally one can adopt different polaron numbers, $n_{p}\geqslant 1$, depending on the approximation one takes.
As a simplest description without loss of accuracy, in
the present work we take $n_{p}=2$ so that $a$ sums over the polaron $\alpha
$ and and antipolaron $\beta $ in the main text.

The variation of the wave function consists of the change resources of the
displacement $x_{a}^{\pm },$ frequency renormalization $\xi _{a}^{\pm },$
polaron weight $c_{a}^{\pm }$ and polaron asymmetry $\delta _{a}^{\pm }$ in
response to the variation of the coupling $g$
\begin{eqnarray}
\frac{d\psi _{\pm }}{dg} &=&\sum\limits_{a}^{n_{p}}\left( c_{a}^{\pm }\frac{%
d\varphi _{a}^{\pm }}{dx_{a}^{\pm }}\frac{dx_{a}^{\pm }}{dg}+c_{a}^{\pm }%
\frac{d\varphi _{a}^{\pm }}{d\xi _{\pm }^{a}}\frac{d\xi _{a}^{\pm }}{dg}%
\right.   \nonumber \\
&&+\left. \frac{dc_{a}^{\pm }}{dg}\varphi _{a}^{\pm }+c_{a}^{\pm }\frac{%
d\varphi _{a}^{\pm }}{d\delta _{a}^{\pm }}\frac{d\delta _{a}^{\pm }}{dg}%
\right) .
\end{eqnarray}%
Then, the QFI can be disassembled into intra-resource parts and
inner-resource parts
\begin{eqnarray}
F_{Q} &=&F_{Q}^{x,x}+F_{Q}^{\xi ,\xi }+F_{Q}^{\delta ,\delta }+F_{Q}^{\rho
,\rho }  \nonumber \\
&&+F_{Q}^{x,\xi }+F_{Q}^{x,\rho }+F_{Q}^{x,\delta }+F_{Q}^{\rho ,\xi
}+F_{Q}^{\rho ,\delta }+F_{Q}^{\xi ,\delta }.
\end{eqnarray}%
The intra-resource parts can be formulated as
\begin{eqnarray}
F_{Q}^{x,x} &=&\sum\limits_{a}^{n_{p}}\sum\limits_{a^{\prime
}}^{n_{p}}\sum\limits_{\sigma =\pm }c_{a}^{\sigma }c_{a^{\prime }}^{\sigma
}\langle \frac{d\varphi _{a}^{\sigma }}{dx_{a}^{\sigma }}|\frac{d\varphi
_{a^{\prime }}^{\sigma }}{dx_{a^{\prime }}^{\sigma }}\rangle \frac{%
dx_{a}^{\sigma }}{dg}\frac{dx_{a^{\prime }}^{\sigma }}{dg}, \\
F_{Q}^{\xi ,\xi } &=&\sum\limits_{a}^{n_{p}}\sum\limits_{a^{\prime
}}^{n_{p}}\sum\limits_{\sigma =\pm }c_{a}^{\sigma }c_{a^{\prime }}^{\sigma
}\langle \frac{d\varphi _{a}^{\sigma }}{d\xi _{a}^{\sigma }}|\frac{d\varphi
_{a^{\prime }}^{\sigma }}{d\xi _{a^{\prime }}^{\sigma }}\rangle \frac{d\xi
_{a}^{\sigma }}{dg}\frac{d\xi _{a^{\prime }}^{\sigma }}{dg}, \\
F_{Q}^{\delta ,\delta } &=&\sum\limits_{a}^{n_{p}}\sum\limits_{a^{\prime
}}^{n_{p}}\sum\limits_{\sigma =\pm }c_{a}^{\sigma }c_{a^{\prime }}^{\sigma
}\langle \frac{d\varphi _{a}^{\sigma }}{d\delta _{a}^{\sigma }}|\frac{%
d\varphi _{a^{\prime }}^{\sigma }}{d\delta _{a^{\prime }}^{\sigma }}\rangle
\frac{d\delta _{a}^{\sigma }}{dg}\frac{d\delta _{a^{\prime }}^{\sigma }}{dg},
\\
F_{Q}^{\rho ,\rho } &=&\sum\limits_{a}^{n_{p}}\sum\limits_{a^{\prime
}}^{n_{p}}\sum\limits_{\sigma =\pm }\langle \varphi _{a}^{\sigma }|\varphi
_{a^{\prime }}^{\sigma }\rangle \frac{dc_{a}^{\sigma }}{dg}\frac{%
dc_{a^{\prime }}^{\sigma }}{dg},
\end{eqnarray}%
while inter-resource parts read
\begin{eqnarray}
F_{Q}^{x,\xi } &=&\sum\limits_{a}^{n_{p}}\sum\limits_{a^{\prime }\neq
a}^{n_{p}}\sum\limits_{\sigma =\pm }\left( c_{a}^{\sigma }c_{a^{\prime
}}^{\sigma }\langle \frac{d\varphi _{a}^{\sigma }}{dx_{a}^{\sigma }}|\frac{%
d\varphi _{a^{\prime }}^{\sigma }}{d\xi _{a^{\prime }}^{\sigma }}\rangle
\frac{dx_{a}^{\sigma }}{dg}\frac{d\xi _{a^{\prime }}^{\sigma }}{dg}\right.
\nonumber \\
&&\left. +c_{a}^{\sigma }c_{a^{\prime }}^{\sigma }\langle \frac{d\varphi
_{a}^{\sigma }}{d\xi _{a}^{\sigma }}|\frac{d\varphi _{a^{\prime }}^{\sigma }%
}{dx_{a^{\prime }}^{\sigma }}\rangle \frac{d\xi _{a}^{\sigma }}{dg}\frac{%
dx_{a^{\prime }}^{\sigma }}{dg}\right) , \\
F_{Q}^{x,\delta } &=&\sum\limits_{a}^{n_{p}}\sum\limits_{a^{\prime }\neq
a}^{n_{p}}\sum\limits_{\sigma =\pm }\left( c_{a}^{\sigma }c_{a^{\prime
}}^{\sigma }\langle \frac{d\varphi _{a}^{\sigma }}{dx_{a}^{\sigma }}|\frac{%
d\varphi _{a^{\prime }}^{\sigma }}{d\delta _{a^{\prime }}^{\sigma }}\rangle
\frac{dx_{a}^{\sigma }}{dg}\frac{d\delta _{a^{\prime }}^{\sigma }}{dg}%
\right.   \nonumber \\
&&\left. +c_{a}^{\sigma }c_{a^{\prime }}^{\sigma }\langle \frac{d\varphi
_{a}^{\sigma }}{d\delta _{a}^{\sigma }}|\frac{d\varphi _{a^{\prime
}}^{\sigma }}{dx_{a^{\prime }}^{\sigma }}\rangle \frac{d\delta _{a}^{\sigma }%
}{dg}\frac{dx_{a^{\prime }}^{\sigma }}{dg}\right) , \\
F_{Q}^{\xi ,\delta } &=&\sum\limits_{a}^{n_{p}}\sum\limits_{a^{\prime }\neq
a}^{n_{p}}\sum\limits_{\sigma =\pm }\left( c_{a}^{\sigma }c_{a^{\prime
}}^{\sigma }\langle \frac{d\varphi _{a}^{\sigma }}{d\xi _{a}^{\sigma }}|%
\frac{d\varphi _{a^{\prime }}^{\sigma }}{d\delta _{a^{\prime }}^{\sigma }}%
\rangle \frac{d\xi _{a}^{\sigma }}{dg}\frac{d\delta _{a^{\prime }}^{\sigma }%
}{dg}\right.   \nonumber \\
&&\left. +c_{a}^{\sigma }c_{a^{\prime }}^{\sigma }\langle \frac{d\varphi
_{a}^{\sigma }}{d\delta _{a}^{\sigma }}|\frac{d\varphi _{a^{\prime
}}^{\sigma }}{d\xi _{a^{\prime }}^{\sigma }}\rangle \frac{d\delta
_{a}^{\sigma }}{dg}\frac{d\xi _{a^{\prime }}^{\sigma }}{dg}\right) , \\
F_{Q}^{\xi ,\rho } &=&\sum\limits_{a}^{n_{p}}\sum\limits_{a^{\prime }\neq
a}^{n_{p}}\sum\limits_{\sigma =\pm }\left( c_{\sigma }^{a}\frac{d\xi
_{a}^{\sigma }}{dg}\langle \frac{d\varphi _{a}^{\sigma }}{d\xi _{a}^{\sigma }%
}|\varphi _{a^{\prime }}^{\sigma }\rangle \frac{dc_{a^{\prime }}^{\sigma }}{%
dg}\right.   \nonumber \\
&&\left. +\frac{dc_{a}^{\sigma }}{dg}\langle \varphi _{a}^{\sigma }|\frac{%
d\varphi _{a^{\prime }}^{\sigma }}{d\xi _{a^{\prime }}^{\sigma }}\rangle
c_{a^{\prime }}^{\sigma }\frac{d\xi _{a^{\prime }}^{\sigma }}{dg}\right) , \\
F_{Q}^{x,\rho } &=&\sum\limits_{a}^{n_{p}}\sum\limits_{a^{\prime }\neq
a}^{n_{p}}\sum\limits_{\sigma =\pm }\left( c_{a}^{\sigma }\frac{%
dx_{a}^{\sigma }}{dg}\langle \frac{d\varphi _{a}^{\sigma }}{dx_{\sigma }}%
|\varphi _{a^{\prime }}^{\sigma }\rangle \frac{dc_{a^{\prime }}^{\sigma }}{dg%
}\right.   \nonumber \\
&&\left. +\frac{dc_{a}^{\sigma }}{dg}\langle \varphi _{a}^{\sigma }|\frac{%
d\varphi _{a^{\prime }}^{\sigma }}{dx_{a^{\prime }}^{\sigma }}\rangle
c_{a^{\prime }}^{\sigma }\frac{dx_{a^{\prime }}^{\sigma }}{dg}\right) ; \\
F_{Q}^{\delta ,\rho } &=&\sum\limits_{a}^{n_{p}}\sum\limits_{a^{\prime }\neq
a}^{n_{p}}\sum\limits_{\sigma =\pm }\left( c_{a}^{\sigma }\frac{%
dx_{a}^{\sigma }}{dg}\langle \frac{d\varphi _{a}^{\sigma }}{dx_{a}^{\sigma }}%
|\varphi _{\sigma }^{a^{\prime }}\rangle \frac{dc_{a^{\prime }}^{\sigma }}{dg%
}\right.   \nonumber \\
&&\left. +\frac{dc_{a}^{\sigma }}{dg}\langle \varphi _{a}^{\sigma }|\frac{%
d\varphi _{a^{\prime }}^{\sigma }}{dx_{\sigma }^{a^{\prime }}}\rangle
c_{a^{\prime }}^{\sigma }\frac{dx_{a^{\prime }}^{\sigma }}{dg}\right) .
\end{eqnarray}%
Here the overlaps of two wave packets $\varphi
_{1},\varphi _{2}$ are defined as the expectation integral
\begin{equation}
\langle \varphi _{1}|\varphi _{2}\rangle =
\int _{-\infty }^{\infty }
\varphi _{1}^{\ast }\left( x\right) \varphi _{2}\left( x\right) dx
\end{equation}%
while the derivative ones are similarly defined by replacing $\varphi _{i}$
with the derivative $d\varphi _{i}/dp_{i}$ with respect to the parameter $%
p_{i}$. The analytic and explicit expressions of the derivative wave-packet overlaps are
too lengthy to list here, nevertheless they can be readily obtained with
Eq.~\eqref{polaron-general}.

\section{Variational expressions in asymmetric polaron picture}
\label{Appendix-Vari-E}

As introduced in Sec.~\ref{Sect-Vari-APP} in the main text, the variational energy is given by
\begin{eqnarray}
	E=\langle\Psi\mid H\mid\Psi\rangle=h_{++}^{+}+\frac{1}{2} P \Omega n_{+-}+\varepsilon_{0}.\label{E-vari}
\end{eqnarray}
The general wave function $\Psi$ is defined at Eq.~\eqref{psi-up-down}, while the form in the asymmetric polaron picture is presented in Eqs.\eqref{eq:ware function}-\eqref{A-factor}. Under such a formalism, the first term in Eq.~\eqref{E-vari}
\begin{eqnarray}
	h_{++}^{+}&&=\langle\psi\mid h^{+}\mid\psi\rangle\notag\\
	&&=\alpha^{2}\langle\varphi_{\alpha}\mid h^{+}\mid\varphi_{\alpha}\rangle
	+\beta^{2}\langle\varphi_{\beta}\mid h^{+}\mid\varphi_{\beta}\rangle\notag\\
	&&+2\alpha\beta\langle\varphi_{\alpha}\mid h^{+}\mid\varphi_{\beta}\rangle,
\end{eqnarray}
gives the effective single-particle energy for $h^{+} =\frac{1}{2}\omega[\hat{p}^{2}+(\hat{x}+g')^2]$ with a constant energy $\varepsilon_{0}=-\frac{1}{2}\omega(g'^{2}+1)$. The middle term in Eq.~\eqref{E-vari} denotes the tunneling or spin-flipping energy with
\begin{eqnarray}
	n_{+-}&&=\langle\psi|\bar{\psi}\rangle\notag\\
	&&=\alpha^{2}\langle\varphi_{\alpha}\mid\bar{\varphi}_{\alpha}\rangle
	+\beta^{2}\langle\varphi_{\beta}\mid\bar{\varphi}_{\beta}\rangle\notag\\
	&&+2\alpha\beta\langle\varphi_{\alpha}\mid\bar{\varphi}_{\beta}\rangle.
 \end{eqnarray}
The expectation factors can be obtained explicitly for intra-polaron term
\begin{eqnarray}
&&\langle\varphi_{\alpha}\mid h^{+}\mid\varphi_{\alpha}\rangle\notag\\
=&&\frac{\omega}{8\xi_{\alpha}^2}\left[{\delta_{\alpha}}^2 \left(2 (\zeta_{\alpha}-1)^2 \xi_{\alpha}{g'}^2 +3\xi_{\alpha}^2+3\right)\right]\notag\\
+&&\frac{\omega}{8\xi_{\alpha}^2}\left[2\xi_{\alpha}\left(2(\zeta_{\alpha}-1)^2\xi_{\alpha}{g'}^2+\xi_{\alpha}^2+1\right)\right]\notag\\
-&&\frac{\omega}{8 \xi_{\alpha}^2}\left[8 \delta_{\alpha}(\zeta_{\alpha}-1)\xi_{\alpha}g'\right],
\end{eqnarray}
intra-antipolaron term
\begin{eqnarray}
&&\langle\varphi_{\beta}\mid h^{+}\mid\varphi_{\beta}\rangle\notag\\
=&&\frac{\omega}{8\xi_{\beta}^2}\left[{\delta_{\beta}}^2 \left(2 (\zeta_{\beta}+1)^2 \xi_{\beta}{g'}^2 +3\xi_{\beta}^2+3\right)\right]\notag\\
+&&\frac{\omega}{8\xi_{\beta}^2}\left[2\xi_{\beta}\left(2(\zeta_{\beta}+1)^2\xi_{\beta}{g'}^2+\xi_{\beta}^2+1\right)\right]\notag\\
+&&\frac{\omega}{8 \xi_{\beta}^2}\left[8 \delta_{\beta}(\zeta_{\beta}+1)\xi_{\beta}g'\right],
\end{eqnarray}
\begin{widetext}
inter-antipolaron term
\begin{eqnarray}
&&\langle\varphi_{\alpha}\mid h^{+}\mid\varphi_{\beta}\rangle
\notag\\
=&& f_1 \left\{
              \xi _{\alpha } \xi _{\beta }\left[
     \xi _{\beta }^2-\delta _{\beta } \left(\zeta _{\beta }+1\right){}^2 \xi _{\beta } \left(g'\right)^3 \left(\zeta _{\alpha }+\zeta _{\beta }\right)+\delta _{\beta } g' \left(-3 \zeta _{\alpha }+\zeta _{\beta }+4\right)-\left(\zeta _{\beta }+1\right) \xi _{\beta } \left(g'\right)^2 \left(2 \zeta _{\alpha }-\zeta _{\beta }-3\right)+2
                                          \right]   \right. \notag\\
     &&\qquad \left. + \xi _{\beta }^2 \left[
               2 \delta _{\beta } \left(\zeta _{\beta }+1\right) g'+\left(\zeta _{\beta }+1\right){}^2 \xi _{\beta } \left(g'\right)^2+1
                                      \right]
      \right\}
\notag\\
+&&  f_1\xi _{\alpha }^3 \left\{
            \xi _{\beta }+\delta _{\beta } \left(g'\right)^3 \left(\zeta _{\alpha }+\zeta _{\beta }\right)
            \left[
              \zeta _{\alpha }^2 \left(\xi _{\beta }^2-1\right)+2 \zeta _{\alpha } \left(\zeta _{\beta } \xi _{\beta }^2+1\right)+\zeta _{\beta }^2 \xi _{\beta }^2-1
            \right]   \right.\notag\\
     &&\qquad \left. - 3 \delta _{\beta } \xi _{\beta } g' \left(\zeta _{\alpha }+\zeta _{\beta }\right)
          -\left(g'\right)^2 \left[
     \zeta _{\alpha }^2 \left(\xi _{\beta }^2-1\right)+2 \zeta _{\alpha } \left(\zeta _{\beta } \xi _{\beta }^2+1\right)+\zeta _{\beta }^2 \xi _{\beta }^2-1
                            \right]
                       \right\}
\notag\\
+&& f_1 \xi _{\alpha }^2 \left\{
     2 \xi _{\beta }^2+2 \left(\zeta _{\alpha }-1\right) \delta _{\beta } \left(\zeta _{\beta }+1\right) \xi _{\beta } \left(g'\right)^3 \left(\zeta _{\alpha }+\zeta _{\beta }\right)-\delta _{\beta } g' \left[ 3 \zeta _{\alpha } \left(\xi _{\beta }^2+1\right)
  +3 \zeta _{\beta } \xi _{\beta }^2+\zeta _{\beta }-2
                                  \right]   \right.\notag\\
 &&\qquad \left.
  -\xi _{\beta } \left(g'\right)^2 \left[
           \zeta _{\alpha }^2 \left(\xi _{\beta }^2-1\right)+2 \zeta _{\alpha } \left(\zeta _{\beta } \xi _{\beta }^2+\zeta _{\beta }+2\right)+\zeta _{\beta }^2 \xi _{\beta }^2-2 \zeta _{\beta }-3      \right]+1
                         \right\}
\notag\\
+&& f_2 \delta _{\beta }\left\{\xi _{\beta }^2 \left[ \left(\zeta _{\beta }+1\right) \xi _{\beta } \left(g'\right)^2 \left(2 \zeta _{\alpha }+3 \zeta _{\beta }+1\right)+3
                                               \right]
\right.\notag\\
  &&\qquad\left. -\xi _{\alpha } \xi _{\beta } \left[
          -3 \left(\xi _{\beta }^2+2\right)+\left(\zeta _{\beta }+1\right){}^2 \xi _{\beta }^2 \left(g'\right)^4 \left(\zeta _{\alpha }+\zeta _{\beta }\right){}^2+3 \left(\zeta _{\alpha }-1\right) \xi _{\beta } \left(g'\right)^2 \left(\zeta _{\alpha }+2 \zeta _{\beta }+1\right)
                                       \right]
   \right\}
   \notag\\
+&& f_2 \delta _{\beta } \xi _{\alpha }^2
    \left\{
6 \xi _{\beta }^2+2 \left(\zeta _{\alpha }-1\right) \left(\zeta _{\beta }+1\right) \xi _{\beta }^2 \left(g'\right)^4 \left(\zeta _{\alpha }+\zeta _{\beta }\right){}^2
\right.\notag\\
    &&\qquad  \left. -3 \xi _{\beta } \left(g'\right)^2
              \left[
    2 \zeta _{\alpha }^2 \xi _{\beta }^2+2 \zeta _{\alpha } \left(2 \zeta _{\beta } \xi _{\beta }^2+\zeta _{\beta }+1\right)+\zeta _{\beta }^2 \left(2 \xi _{\beta }^2+1\right)-1
              \right]+3
    \right\}
\notag\\
+&& f_2 \delta _{\beta } \xi _{\alpha }^3
       \left\{
3 \xi _{\beta }+\xi _{\beta } \left(g'\right)^4 \left(\zeta _{\alpha }+\zeta _{\beta }\right){}^2
          \left[
\zeta _{\alpha }^2 \left(\xi _{\beta }^2-1\right)+2 \zeta _{\alpha } \left(\zeta _{\beta } \xi _{\beta }^2+1\right)+\zeta _{\beta }^2 \xi _{\beta }^2-1
          \right]
\right.\notag\\
    &&\qquad \left. -\left(g'\right)^2
    \left[
    \zeta _{\alpha }^2 \left(6 \xi _{\beta }^2-3\right)+2 \zeta _{\alpha }\zeta _{\beta } \left(6 \xi _{\beta }^2-1\right)+4\zeta _{\alpha }+6 \zeta _{\beta }^2 \xi _{\beta }^2+2 \zeta _{\beta }-1
    \right]
       \right\},
\end{eqnarray}
\end{widetext}
and inter-spin terms
\begin{eqnarray}
&&\langle\varphi_{\alpha}\mid\bar{\varphi}_{\alpha}\rangle=
\frac{e^{-\zeta _{\alpha }^2 \xi _{\alpha }\left(g'\right)^2}}{2 \xi _{\alpha }}
   \left[2 \xi _{\alpha } \left(\delta _{\alpha } \zeta _{\alpha } g'+1\right){}^2-\delta _{\alpha }^2
   \right],\\
&&\langle\varphi_{\beta}\mid\bar{\varphi}_{\beta}\rangle=
\frac{e^{-\zeta _{\beta }^2 \xi _{\beta } \left(g'\right)^2}}{2 \xi _{\beta }}
   \left[2 \xi _{\beta } \left(\delta _{\beta } \zeta _{\beta } g'-1\right){}^2-\delta _{\beta }^2
   \right],\\
&&\langle\varphi_{\alpha}\mid\bar{\varphi}_{\beta}\rangle=
f_3
\left[\xi _{\alpha }+\xi _{\beta }+\xi _{\alpha } \delta _{\beta } g' \left(\zeta _{\alpha }-\zeta _{\beta }\right)\right]
\notag\\
&&\quad +f_4
    \left\{
\xi _{\alpha } \left(-\delta _{\beta }\right)+\delta _{\beta } \xi _{\beta }
                 \left[
\xi _{\alpha } \left(g'\right)^2 \left(\zeta _{\alpha }-\zeta _{\beta }\right){}^2-1
                 \right] \right. \notag\\
&&\qquad \left. +\xi _{\beta } g' \left(\zeta _{\alpha }-\zeta _{\beta }\right) \left(\xi _{\alpha }+\xi _{\beta }\right)
    \right\},
\end{eqnarray}
where the coefficients are give by
\begin{eqnarray}
&&f_1=\frac{\omega\sqrt[4]{\xi _{\alpha }} \sqrt[4]{\xi _{\beta }}}{\sqrt{2} \left(\xi _{\alpha }+\xi _{\beta }\right)^{7/2}}\exp
\left[
\frac{\xi _{\alpha } \xi _{\beta } \left(g'\right)^2 \left(\zeta _{\alpha }+\zeta _{\beta }\right){}^2}
     {-2 \left(\xi _{\alpha }+\xi _{\beta }\right)}
\right],
\\
&&f_2=\frac{\omega  \delta _{\alpha } \sqrt[4]{\xi _{\alpha }} \sqrt[4]{\xi _{\beta }}}{\sqrt{2} \left(\xi _{\alpha }+\xi _{\beta }\right){}^{9/2}}\exp
\left[
\frac{\xi _{\alpha } \xi _{\beta } \left(g'\right)^2 \left(\zeta _{\alpha }+\zeta _{\beta }\right){}^2}
      {-2 \left(\xi _{\alpha }+\xi _{\beta }\right)}
\right],
\\
&&f_3=\frac{\sqrt{2} \sqrt[4]{\xi _{\alpha }} \sqrt[4]{\xi _{\beta }} \exp
\left[
-\frac{\xi _{\alpha } \xi _{\beta } \left(g'\right)^2 \left(\zeta _{\alpha }-\zeta _{\beta }\right){}^2}{2 \left(\xi _{\alpha }+\xi _{\beta }\right)}
\right]
     }
      {\left(\xi _{\alpha }+\xi _{\beta }\right){}^{3/2}},
\\
&&f_4=
\frac{\sqrt{2} \delta _{\alpha } \sqrt[4]{\xi _{\alpha }} \sqrt[4]{\xi _{\beta }} \exp
\left[
-\frac{\xi _{\alpha } \xi _{\beta } \left(g'\right)^2 \left(\zeta _{\alpha }-\zeta _{\beta }\right){}^2}{2 \left(\xi _{\alpha }+\xi _{\beta }\right)}
\right]
     }
{\left(\xi _{\alpha }+\xi _{\beta }\right){}^{5/2}}.
\end{eqnarray}

\bibliography{Refs-2025-12-Asym-polaron}

\end{document}